\documentclass[preprint,12pt]{elsarticle} 

\usepackage{amssymb}

\newcommand{\qt}[1]{``#1''}
\newcommand{\qo}[1]{`#1'}

\usepackage{tablefootnote}
\usepackage{makecell}
\usepackage{hyperref}
\usepackage{subcaption}
\usepackage{comment}
\usepackage{adjustbox}
\usepackage{color}
\usepackage{graphicx}
\usepackage{booktabs}
\usepackage{cleveref}
\usepackage{colortbl}
\usepackage{xfrac}
\usepackage{multirow}
\usepackage[table, svgnames]{xcolor}

\journal{Journal}

\begin{document}

\begin{frontmatter}



\title{Do you feel safe with your robot? Factors Influencing Perceived Safety in Human-Robot Interaction based on Subjective and Objective Measures}


\author[inst1]{Neziha Akalin}

\affiliation[inst1]{organization={School of Science and Technology, \"{O}rebro University},
            city={\"{O}rebro},
            postcode={SE-701 82}, 
            country={Sweden}}

\author[inst2]{Annica Kristoffersson}

\author[inst1]{Amy Loutfi}

\affiliation[inst2]{organization={School of Innovation, Design and Engineering, M\"{a}lardalen University},
            city={V\"{a}ster\r{a}s},
            postcode={SE-721 23}, 
            country={Sweden}}

\begin{abstract}

Safety in human-robot interaction can be divided into physical safety and perceived safety, where the latter is still under-addressed in the literature. Investigating perceived safety in human-robot interaction requires a multidisciplinary perspective. Indeed, perceived safety is often considered as being associated with several common factors studied in other disciplines, i.e., comfort, predictability, sense of control, and trust. In this paper, we investigated the relationship between these factors and perceived safety in human-robot interaction using subjective and objective measures. We conducted a two-by-five mixed-subjects design experiment. There were two between-subjects conditions: the faulty robot was experienced at the beginning or the end of the interaction. The five within-subjects conditions correspond to (1) baseline, and the manipulations of robot behaviors to stimulate: (2) discomfort, (3) decreased perceived safety, (4) decreased sense of control and (5) distrust. The idea of triggering a deprivation of these factors was motivated by the definition of safety in the literature where safety is often defined by the absence of it. Twenty-seven young adult participants took part in the experiments. Participants were asked to answer questionnaires that measure the manipulated factors after within-subjects conditions. Besides questionnaire data, we collected objective measures such as videos and physiological data. The questionnaire results show a correlation between comfort, sense of control, trust, and perceived safety. Since these factors are the main factors that influence perceived safety, they should be considered in human-robot interaction design decisions. We also discuss the effect of individual human characteristics (such as personality and gender) that they could be predictors of perceived safety. We used the physiological signal data and facial affect from videos for estimating  perceived safety where participants' subjective ratings were utilized as labels. The data from objective measures revealed that the prediction rate was higher from physiological signal data. This paper can play an important role in the goal of better understanding perceived safety in human-robot interaction.

\end{abstract} 


\end{frontmatter}


\section{Introduction}
\label{sec:intro}
Safety is an essential property of daily life given its critical role in being one of the fundamental needs of human beings~\cite{Maslow1943}. Since robotic systems should be designed without compromising human safety, there is a plethora of research on physical safety in human-robot interaction (HRI). The physical safety in HRI has been implemented in many different ways, including human–robot collaborative control schemes~\cite{su2019improved}, deep learning approaches~\cite{su2020deep}, and teaching by demonstration~\cite{su2021toward}. A robot that is designed to coexist with humans must be safe not only concerning from causing physical harm but from causing psychological harm. Still, the safety perception of the users has been a tendency to be overlooked both in HRI literature and in safety standards~\cite{salvini2021safety}. Yet, perceived safety is crucial for long-term interaction, collaboration, and acceptance. For acceptable HRI, a robot must avoid taking actions that might cause fear, surprise, discomfort or create an unpleasant social situation for humans even if its actions do not cause any physical harm~\cite{sisbot2010synthesizing}. Indeed, there may even be a discrepancy between physical safety and safety perception~\cite{salem2015towards}, and it has been shown that maintenance of physical safety by simply preventing collisions can still lead to a lower degrees of perceived safety~\cite{lasota2015analyzing}. 

The challenge of assessing perceived safety is further compounded for a special class of robots, namely, domestic and social robots. While in industry, robot operators receive professional training before they interact with robots, anyone could potentially interact with domestic robots without receiving any training. Moreover, social robots expected to serve in domestic environments may used by vulnerable users, such as older adults and children. In this respect, perceived safety during HRI deserves much attention considering the psychological, cognitive, and emotional consequences of interactions. Said differently, \textit{How can we design and implement systems such that the users perceive them safe to interact with?}

\begin{figure}[!htbp]
\begin{center}
\includegraphics[width=0.60\textwidth]{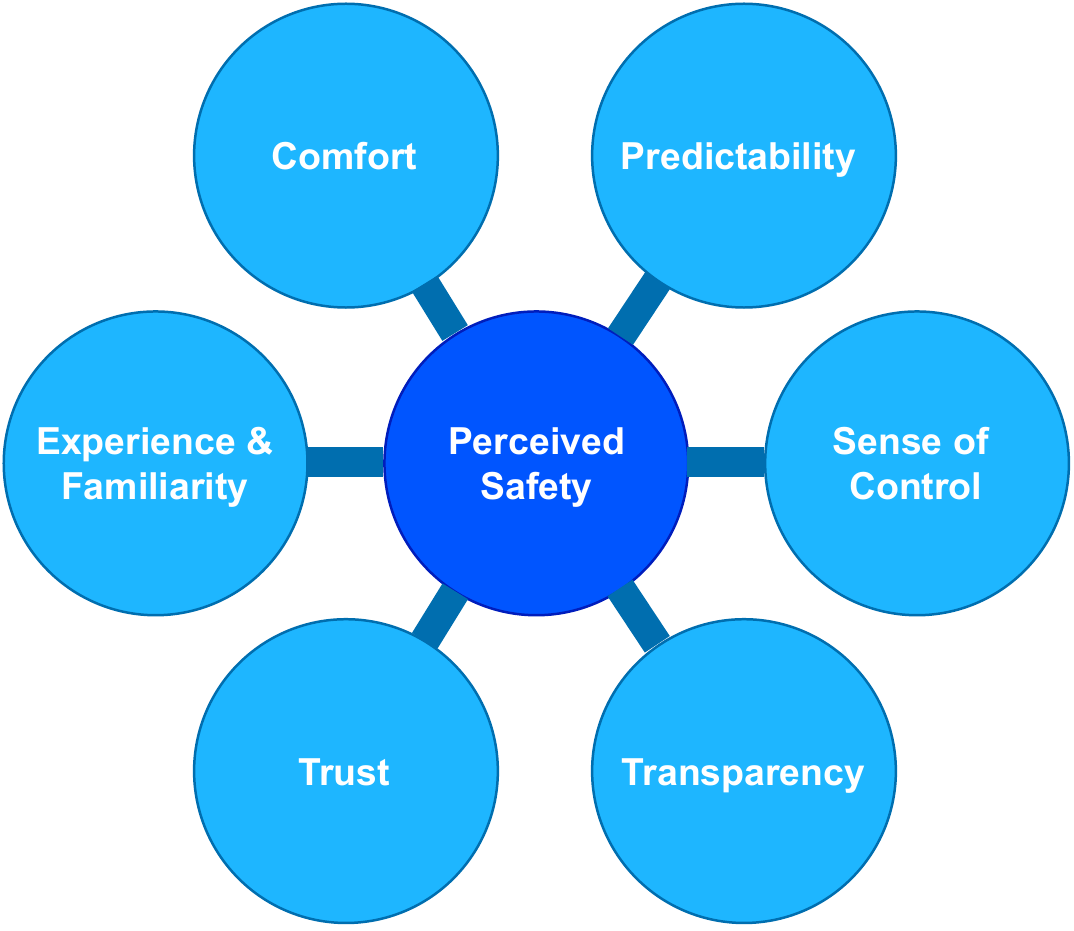}
\caption{The factors influencing perceived safety.}
\label{fig:perceived_safety}       
\end{center}
\end{figure}

Before we can answer this question, we first need to come to an understanding of what perceived safety is. In our previous work, we investigated the sense of safety and security of older people during HRI~\cite{nezihaChapter}. The term~\qt{sense of safety and security} was borrowed from the gerontology literature. Moreover, we reported the factors influencing the sense of safety and security by consulting gerontology literature, HRI literature, and our user studies~\cite{nezihaChapter, akalin2017evaluation}. In this current study, as a starting point, we reviewed the multidisciplinary perspectives of perceived safety. It showed that the factors identified in our previous work~\cite{nezihaChapter} align with perceived safety of general user profiles. While each discipline views perceived safety from its unique perspective, there are several common factors associated with perceived safety: comfort, predictable situations, familiar situations (having experience), sense of control, and trust (Figure~\ref{fig:perceived_safety}).

Building on~\cite{nezihaChapter}, this paper provides a step further with a user study to explore the relationships between perceived safety and the factors mentioned above. We devised a two-by-five mixed-subjects design experiment. The two between-subjects conditions were the faulty robot was experienced at the beginning or at the end of the interaction. We designed these conditions to explore the impact that establishing trust at the beginning of the interaction has on perceived safety. The five within-subjects conditions were (1) baseline, and the manipulations of robot behaviors to stimulate: (2) discomfort, (3) decreased perceived safety, (4) decreased sense of control, and (5) distrust. These manipulations were motivated by the argument that there is nothing to measure in the presence of safety~\cite{hollnagel2014safety}. Therefore, the conditions aimed to stimulate decreased perceived safety. Twenty-seven young adult participants took part in the user study. In the experiments, we collected data through questionnaires, videos, and physiological signals.

The experimental results have shown that individual human characteristics, such as gender and personality traits, influence perceived safety of humans in HRI. People with low neurotic personality traits felt safer and more in control during the interaction. Male participants felt safer, more in control, and more positive than female participants throughout the interaction. The faulty robot being used at the beginning of the interaction or at the end of the interaction did not influence perceived safety or the other factors. As expected, all subjective ratings were highest at the baseline condition. The subjective ratings showed that our manipulations on the robot behaviors for creating discomfort, decreased sense of control, and distrust were successful. These manipulations influenced perceived safety of participants. Short-term unpredictable robot behaviors that did not affect the main functionality of the robot did not influence perceived safety or the other factors. The results showed that perceived safety is correlated with comfort, sense of control, and trust. Moreover, there were also varying degrees of correlation between other factors. To exemplify, there was a strong positive correlation between comfort and sense of control ratings, and a moderate positive correlation between trust and sense of control ratings. This suggests that when participants felt in control over the interaction, they were also comfortable and trusted the robot. When we used subjective perceived safety ratings as labels and classified facial emotions and physiological data, the prediction rate on physiological data was higher.

The paper is organized as follows: Section~\ref{sec:perceived_safety}, we first discuss perceived safety in HRI and from a multidisciplinary perspective, then the section continues with the key factors influencing perceived safety. Section~\ref{sec:experiment} explains the user study. Section~\ref{sec:results} presents the experimental results. In Section~\ref{sec:discussion}, we provide a discussion regarding the implications of the experimental results, the limitations and future research directions. Finally, Section~\ref{sec:conclusion} concludes the paper.

\section{Perceived Safety}
\label{sec:perceived_safety}

The nomenclature for perceived safety varies in different disciplines and application areas. As an example, the term \qt{psychological safety} is used in work environment safety~\cite{kahn1990psychological}, team and group dynamics studies~\cite{edmondson2004psychological}. \qt{Sense of safety and security} is used in gerontology literature~\cite{fonad2006moving} and \qt{perceived safety} is used in various disciplines~\cite{raue2019perceived}. Similarly, HRI literature adopted several different terms for the safety perception: psychological safety~\cite{lasota2017survey, kamide2012new}, sense of safety and security~\cite{nezihaChapter}, perceived safety~\cite{bartneck2009measurement}, mental safety~\cite{matsas2017design}, and sense of security~\cite{nonaka2004evaluation, nyholm2021users}.  

Lasota et al.~\cite{lasota2017survey} presented a survey of potential methods enabling safe HRI. This work considered psychological safety in the context of HRI as interactions that are stress-free and comfortable. Moreover, to maintain psychological safety, it should be ensured that the robot's motion, appearance, embodiment, gaze, speech, posture, social conduct, or any other attribute do not result in any psychological discomfort or stress~\cite{lasota2017survey}. In~\cite{bartneck2009measurement}, Bartneck et al. proposed a questionnaire series called Godspeed Questionnaire to measure anthropomorphism, animacy, likeability, perceived intelligence, and perceived safety. In the same study, authors defined perceived safety as \qt{the user's perception of the level of danger when interacting with a robot, and the user's level of comfort during the interaction} (p.~76). Lichtenthaler et al.~\cite{lichtenthaler2012influence} focused on the influence of legibility on perceived safety in situations where a robot crosses a human's path. The legible robot behaviors were explained as behaviors in which the next actions are predictable and behaviors that carry out the expectations of a human interactant. They reported that there was a correlation between perceived safety and legibility. Moreover, legible robot behaviors resulted in higher perceived safety. 

Matsas et al.~\cite{matsas2017design} presented a virtual reality training system for human-robot collaboration in industrial settings where the definition for mental safety is given as~\qt{the enhanced users' vigilance and awareness of the robot motion, that will not cause any unpleasantness such as fear, shock or surprise.} (p.~140). Nanoka et al.~\cite{nonaka2004evaluation} conducted experiments to evaluate the participants' sense of security. In the experiments, the virtual robots in varying shape, size and motions were presented to participants in a pick and place scenario. They observed that robots' human-like behaviors made the humans feel more comfortable. Nyholm et al.~\cite{nyholm2021users} discussed users' sense of security with humanoid robots in the healthcare context. They first showed a video of the Pepper robot to participants, and following the video, they conducted semi-structured interviews with 12 participants from different professional groups. The study revealed that participants had ambivalent feelings about robots, as such they perceived humanoid robots to be reliable and unreliable, safe and unsafe, likable and scary, caring and uncaring. 

\subsection{Perceived Safety in Multidisciplinary Context}
\label{subsec:multidisciplinary}

Perceived safety is a term, which is commonly used in different fields including tourism~\cite{rittichainuwat2013tourists}, healthcare services~\cite{bradshaw2014measuring}, urban and environmental studies~\cite{ramirez2021measuring}, clinical psychology~\cite{brosschot2016default}, robotics~\cite{bartneck2009measurement}, and autonomous systems~\cite{XU2018320, kong2018effects}. However, perceived safety is not limited to these fields, a basic search in Web of Science results in more than one hundred categories. When reading articles from several disciplines, we observed that it is common to describe perceived safety with positive affective states such as relieved, comfortable, and assured or lack of perceived safety with negative affective states such as stress, discomfort, fear, and anxiety. For example, in a tourism study, Rittichainuwat~\cite{rittichainuwat2013tourists} explained the safety concern, as an affective experience that is an overlapping emotion of worry, fear, and anxiety that emerges from a nervous situation. In a similar vein, a clinical psychology study, Brosschot et al.~\cite{brosschot2016default} stated that the lack of perceived safety triggers chronic anxiety and stress. For living organisms, unpredictable and uncertain situations are always perceived as unsafe even if there is no threat~\cite{brosschot2016default}. A recent study of urban space safety~\cite{ramirez2021measuring} reported that in a survey of perception of public spaces, characteristics of the respondents such as gender, mobility pattern, and income affected their perceived safety.

Automated vehicles (AVs) is another research area in which perceived safety has received attention in recent years. Similar to HRI, perceived safety of new technologies like AVs are crucial for their acceptance. For example, Xu et al.~\cite{XU2018320} examined the influence of trust and perceived safety on AVs' acceptance and intention to use. The authors defined perceived safety as~\qt{a climate in which drivers and passengers can feel relaxed, safe, and comfortable while driving} (p.~323). Their findings showed that perceived usefulness, trust, and perceived safety were direct predictors of acceptance of AVs. Another study by Moody et al.~\cite{moody2020public} revealed that an individual's socio-demographic characteristics (such as age, gender, education level, employment, and income) and awareness of AVs technology are important factors for perceived safety.

A recent book handled the multidisciplinary perspective of perceived safety~\cite{raue2019perceived}. In the book, safety is considered as a value of a function that includes some degree of distress (occurs in unsafe conditions) and relaxation (occurs in safe conditions) and ranges from \qo{danger} to \qo{peace of mind}~\cite{proske2019safety}. From the psychological point of view, many different components of human life can affect perceived safety including current health status, experienced exposure to crime, financial situation, and social relationships~\cite{eller2019psychological}. The book provides a discussion about the factors that can influence perceived safety~\cite{raue2019perceived}. For example, other people's behaviors and expressions, as well as being accepted and approved by others are fundamental conditions for perceived safety~\cite{raue2019perceived}. Additional important factors are feeling treated fairly and respectfully, having an impact on a given situation, anticipating ongoing events, having certain freedom in what to do, and how to do~\cite{raue2019perceived}. While the situations that are unpredictable or unclear are perceived as unsafe~\cite{kahn1990psychological}, transparent and constructive feedback could promote perceived safety~\cite{raue2019perceived}.

\subsection{Evaluating Perceived Safety in HRI}
\label{subsec:evaluation}

The survey in~\cite{lasota2017survey} touched upon the psychological safety aspects, and the assessment methods for psychological safety during HRI. These methods include physiological sensing, questionnaires, and behavioral metrics. One example of physiological sensing can be seen in Nonaka et al.~\cite{nonaka2004evaluation}, they collected the heart rate of participants, but they reported that there was no relationship between heart rate and the human sense of security in their experimental data. 

Bartneck et al.~\cite{bartneck2009measurement} presented a semantic differential questionnaire as a measurement instrument for perceived safety of robots. Kamide et al.~\cite{kamide2012new} presented a questionnaire for measuring the psychological safety of humanoids. In~\cite{nonaka2004evaluation}, participants rated their emotions using a questionnaire including the items surprise, fear, disgust, and unpleasantness changing between 1 (never) and 6 (very much). In our previous work~\cite{nezihaChapter}, the participants rated their safety perception in a semantic differential questionnaire. In this study, we used all three kinds of methods mentioned in~\cite{lasota2017survey}, namely questionnaires, physiological sensing, and facial affect metrics of the participants for evaluating perceived safety. 

\subsection{Factors Influencing Perceived Safety}
\label{subsec:factors}

Modeling perceived safety  is a challenging task since personal, social, and interpersonal factors can affect it. In addition, in the context of HRI, the robot's properties such as its appearance (embodiment, size, shape, posture, etc.), and its motion (speed, acceleration, proximity to the human, etc.) influence perceived safety. As an example~\cite{haring2016people} reported that an android robot was perceived significantly less safe in comparison to a humanoid and non-biomimetic robot (Keepon robot). Despite the fact that various terms are available for safety perception in different disciplines, we observed that feeling of safety is commonly considered to be related to the same factors such as trust~\cite{kahn1990psychological, edmondson2004psychological, raue2019perceived, proske2008catalogue}, comfort~\cite{kahn1990psychological, edmondson2004psychological}, sense of control~\cite{proske2008catalogue, cao2021development}, experience and familiarity~\cite{raue2019perceived, proske2008catalogue, cao2021development}~\, and uncertainty and predictability~\cite{raue2019perceived, lichtenthaler2012influence,  brosschot2016default, proske2008catalogue, cao2021development}. 

We relate the uncertainty to the sense of security in the sense of safety and security model~\cite{nezihaChapter}. All the other factors match with our previous work where the human-related components of sense of safety and security in older people-robot interaction were defined as comfort, experience, sense of security, sense of control, and trust. These factors cover the key referents of perceived safety in the literature of several disciplines. Although they do not correspond exactly to each discipline's view, they do capture significant factors of perceived safety. Due to the bidirectional nature of the HRI, human-related and robot-related factors cannot be treated separately from each other. For example, gestures of the robot may lead to discomfort in the human, or the software failure of the robot may lead to distrust in the human. 

After analyzing perceived safety from several perspectives, we provide the following definition: perceived safety refers that \textit{the consequences of robot-related factors~\cite{nezihaChapter} (i.e., physical, functional, social properties and gestures of a robot) do not cause distrust, discomfort, lack of control over the interaction; and the person feels familiar with the robot and the situations that are the results of the robot's behaviors. The person feels confident and safe in what actions the robot takes and why the robot takes those actions.} 

To better understand perceived safety in HRI, it is necessary to investigate perceived safety from a different point of view by going beyond the thematic limitation. Since HRI includes two parties (i.e., humans and robots), safety perception is never based on the robot properties alone. We compiled the discussed multidisciplinary perspective of perceived safety in a user study in which we observed the effects of selected robot-related factors on human-related factors. Besides these factors, we examined the users' affective states as in~\cite{raue2019perceived, rittichainuwat2013tourists, brosschot2016default}, and individual human characteristics as analyzed in~\cite{raue2019perceived, ramirez2021measuring, moody2020public}. 

\section{Experimental Design}
\label{sec:experiment}

To investigate the multidisciplinary perspective of perceived safety in HRI, we conducted a two-by-five mixed-subject design experiment with 27 participants. The experimental scenario consisted of playing a quiz game with a robot. The between-subjects conditions were the faulty robot was experienced at the beginning or at the end of the interaction. For within-subjects conditions, we manipulated one factor at a time (comfort, predictability, sense of control, and trust). To decide how to manipulate these factors, we consulted both the HRI and the multidisciplinary literature. The five within-subjects conditions were (1) baseline, and the manipulations of robot behaviors to stimulate: (2) discomfort, (3) decreased perceived safety, (4) decreased sense of control, and (5) distrust. These conditions are described in Section~\ref{sec:exp_conds}.

Throughout the experiments, we collected questionnaires, videos, and physiological signals. The questionnaire data were analyzed to understand how the influencing factors of perceived safety relate to each other and whether any of them have a larger effect on perceived safety. Moreover, we analyzed the effect of personal characteristics (personality and gender) on perceived safety. Additionally, participants' facial and physiological reactions were examined.

\subsection{Research Questions}
\label{subsec:RQ}

We addressed the following research questions in this study:

\begin{itemize}
\item \textit{RQ 1:} What are the relationships between individual human characteristics (i.e., gender, personality traits) and perceived safety during HRI? 
\item \textit{RQ 2:} What is the effect of a faulty robot being at the beginning and the end of the interaction on perceived safety and the influencing factors? 
\item \textit{RQ 3:} How do manipulations of each factor influence the comfort, sense of control, perceived safety and trust of the participants? 
\item \textit{RQ 4:} What is the relationship between perceived safety and the other factors (comfort, sense of control, and trust)?
\item \textit{RQ 5:} Can we predict perceived safety from facial affect and physiological signals?
\end{itemize}

\subsection{Participants} 
\label{subsec:participants}

Twenty-seven participants, 10 males and 17 females ranging from 20 to 37 years of age ($M = 26.51$, $SD = 4.49$) took part in the experiment. Participants were recruited using social media platforms and flyers. We had two different between-subjects experimental setups: \textit{SetupA} and \textit{SetupB}. \textit{SetupA} had 14 participants (8 females) with an average age of 26.35 years ($SD  = 4.55$), and \textit{Setup B} had 13 participants (9 females) with an average age of 26.69 years ($SD = 4.60$). Participants were mostly university students with a non-technical background (law, music, health sciences, social sciences, etc.) from different levels (undergraduate and graduate). When asked about the participants' experience with robots, most of them were unfamiliar with robots. Only one participant had interacted with a robot. A total of eight participants had seen a robot before but not interacted with one. Three of them had seen the Pepper robot from a distance in one of the university activities. A majority of the participants (18 persons) had not seen a real robot prior to the experiment.

\subsection{The Robot and the Game}
\label{subsec:robot_game}

The robot used in our study was the Pepper robot~\cite{pandey2018mass}. The Pepper is a social humanoid robot that supports two-way communication using natural language through a text-to-speech software. It has a curvy design that is friendly looking and engaging. The robot's face is static but its 20 degrees of freedom allows it to gesture with simple body language. It has a height of 120 cm. There is a 10.1-inch touchscreen on the robot's chest. The tablet and LED lights around its eyes can be used to support spoken communication.

The experimental scenario was to play a quiz game with the Pepper robot. The speech was the primary driver of the interaction whereas the robot's tablet was used to support the interaction. The questions were asked by the robot using speech synthesis, and four choices were shown on the tablet. The questions and robot behaviors were scripted, and the robot's speech recognition was controlled by a Wizard-of-Oz method. The participant answered the questions by speech. After the participant answered each question, immediate feedback on whether the answer was correct or incorrect was provided by the robot. The robot's scripted actions were programmed using Python and an SDK called NAOqi provided by Softbank Robotics. The robot gestured while talking, these gestures were the default gestures that come with the text-to-speech module of the SDK. 

The quiz game included 30 general knowledge questions from different categories such as movies, books, countries, information technology, and simple arithmetic operations. The participants were informed that they would play 20 questions, however, they could end the game whenever they wanted after 20 questions. If they decided to finish the game at any moment (after 20 questions), then they would get all the collected points and win the game. However, if they did not finish the game until the last question, then they would share total points with the robot. The role of the robot was introduced as a presenter, teammate, and opponent in this quiz game scenario. The robot was a presenter who asked questions with speech and showed the options on its tablet. The robot was a teammate who could answer six questions (out of 30 questions) if the participant wanted the robot to do so. The robot was an opponent who could finish the game and win the game by getting all the collected points.  However, the robot was not programmed to finish the game. 

Questions were randomly selected in each session from the question set that included 60 questions. After a round of four questions had been asked and answered, the robot approached the participant and the participant filled out questionnaires using the touchscreen tablet on the robot's chest. These between-conditions questionnaires included four parts: comfort questionnaire (Section~\ref{subsec:comfort_questionnaire}), perceived safety questionnaire (Section~\ref{subsec:ps_questionnaire}), sense of control questionnaire (Section~\ref{subsec:scontrol_questionnaire}), and trust questionnaire (Section~\ref{subsec:trust_questionnaire}). While the participant filled out the questionnaire, the robot stood still without showing any lifelike body movements.

\subsection{Experimental Procedure} 
\label{subsec:procedure}
When a participant arrived to the experiment room, the experimenter explained the study as concerned with \qo{how we can make interactions better with social robots}. The participants were informed about the experiment procedure but not about the different conditions that they would encounter during the interaction. Thereafter, participants read and signed the informed consent form. The form included two parts: general information about the study procedure and the consent certificate (information concerning data privacy and consent to record on video). The experiment, the informed consent form and the administered questionnaires were in English. Participants received a lunch coupon (around 8 Euros) as compensation for their participation. This research was approved by the Swedish ethics committee for studies involving human participants. 

\begin{figure}[!htbp]
\centering
        \begin{subfigure}[b]{0.475\textwidth}
        \centering
        \subfloat[\centering Experiment setup.\label{subfig:exp_setup}]{{\includegraphics[width=\textwidth]{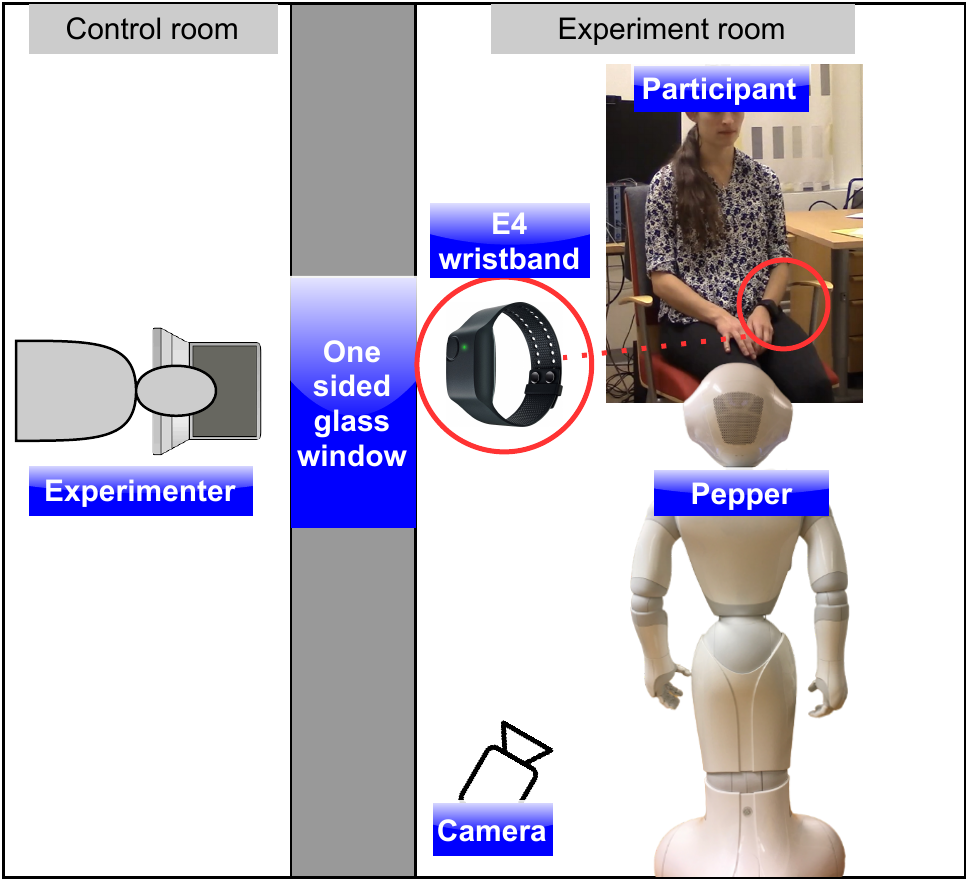} }}
        {{\small}}    
        \end{subfigure}
        \hfill
        \begin{subfigure}[b]{0.475\textwidth}  
            \centering 
            \subfloat[\centering A participant fills out the questionnaire.\label{subfig:participant}]{{\includegraphics[width=\textwidth]{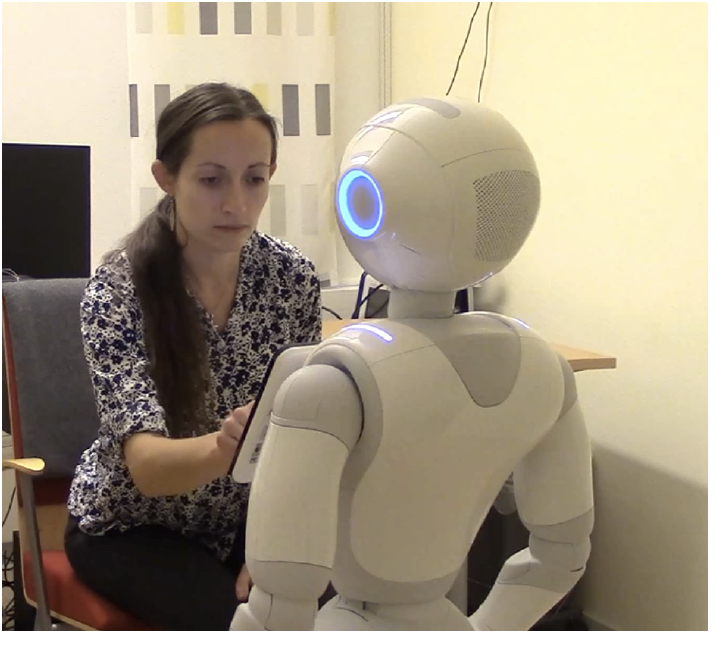} }}
            {{\small}}    
        \end{subfigure}
\caption{A topological overview of the experimental setup in (a), and an image in which the participant (the image used with consent) fills out the questionnaire in (b). The image in (b) was taken by the camera that was positioned as shown in (a).}
\end{figure}

The study was setup in a room which was equipped with a camera, E4 wristband, the robot Pepper, and a chair for the participant. There was an adjacent room which was used as a control room, the two rooms were split by a wall and one-sided mirror glass. The control room was used by the experimenter where she observed the experiment and controlled the speech of the robot. The topological overview of the  experimental setup is given in Figure~\ref{subfig:exp_setup}.
Participants were asked to sit throughout the interaction. The experimenter only interrupted if there were any problems with the robot. The five within-subject conditions in the experiment were: %
\begin{itemize}
    \item Baseline (C1)
    \item Comfort manipulation (C2)
    \item Unpredictable robot behaviors (C3) 
    \item Sense of control manipulation (C4) 
    \item Trust manipulation (C5)
\end{itemize}

\begin{figure}[!htbp]
\begin{center}
\includegraphics[width=1\textwidth]{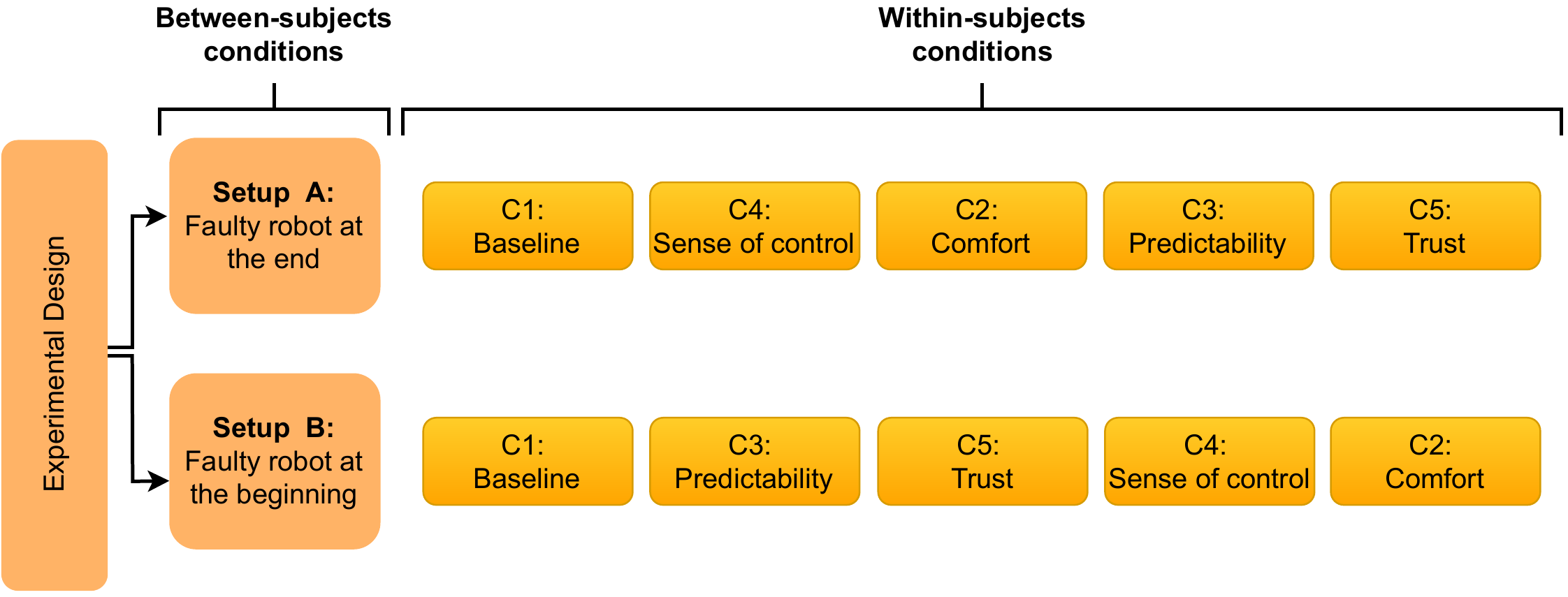}
\caption{Experimental design. There were two between-subjects conditions and five within-subjects conditions. Participants in SetupA and SetupB experienced the five conditions in different order.} 
\label{fig:setupss}     
\end{center}
\end{figure}

These conditions were ordered in two different ways which we call \textit{SetupA} and \textit{SetupB}. The only difference between these two was the order of the conditions. In \textit{SetupA}, C3 and C5 occurred after the other conditions, the order was as follows: C1, C4, C2, C3, and C5. In \textit{Setup B}, C3 and C5 occurred after the baseline condition, the order was as follows: C1, C3, C5, C4, and C2. The experimental design is given in Figure~\ref{fig:setupss}.
The rationale behind these setups was to explore the impact that establishing trust at the beginning of the interaction has on perceived safety. Just as the first impression is important in making a judgment about someone in human-human interaction, it is also important in HRI. As an example, Yu et al.~\cite{yu2017user} showed that participants formed their subjective perceptions of trust in the early stages of the interaction and then adjusted them based on the performance of the system.
Table~\ref{tab:related} shows the human-related and robot-related factors in these conditions. 
\begin{table}[!htbp]
\centering
\caption{Human-related and robot-related factors in experimental conditions. Each condition was designed to manipulate a human-related factor through robot-related factors. The purpose here was to manipulate mainly the corresponding factor (e.g., trust), that manipulation might also affect other factors (e.g., comfort, sense of control). C1 is the baseline that does not have any manipulation. C1 was designed to familiarize the user with the baseline behavior of the robot.}
\label{tab:related} 
\begin{tabular}{lll}\toprule
Condition & Robot-related factor & Human-related factor \\
\midrule
\rowcolor{Lavender!80!gray}  C1 & - & - \\
C2 & Robot's feedback & Comfort \\
\rowcolor{Lavender!80!gray} C3 & Unpredictable behaviors & Perceived safety \\
C4 & Persistent utterances & Sense of control \\
\rowcolor{Lavender!80!gray} C5 & System failure & Trust \\
\bottomrule
\end{tabular}
\end{table}

\begin{figure}[!ht]
\begin{center}
\includegraphics[width=1\textwidth]{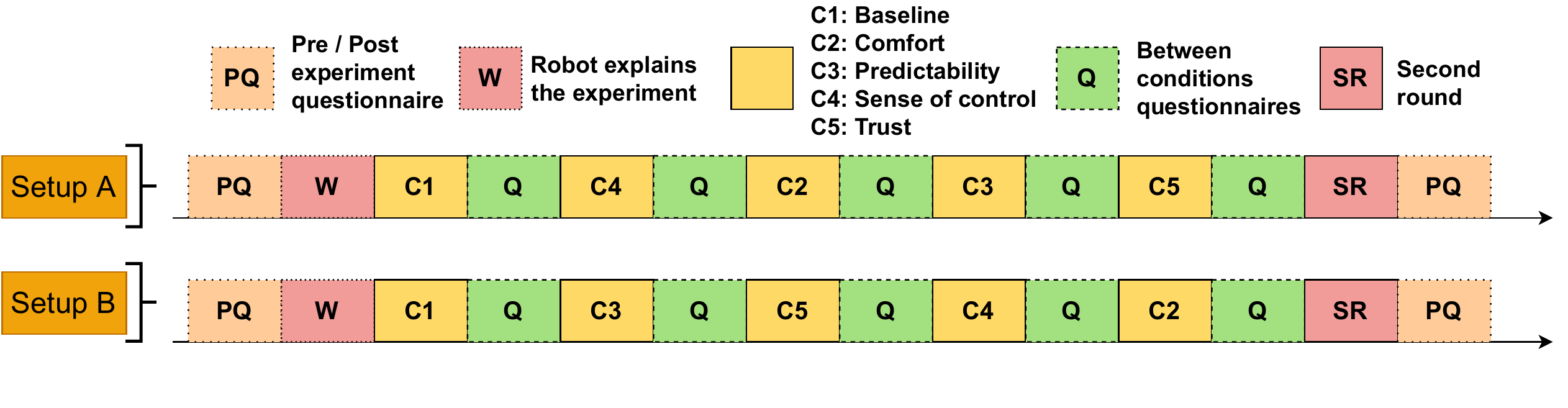}
\caption{Interaction timeline of the game for SetupA and SetupB. Each interaction began with a pre-experiment questionnaire. Interaction proceeded with the robot's introduction of the game. Then, the game started with the baseline condition. Participants filled out between-conditions questionnaires after each condition.} 
\label{fig:flow}     
\end{center}
\end{figure}

Participants were randomly assigned to one of the two setups. After each condition (i.e., C1 - C5) participants filled out questionnaires. Interaction timeline of the experiment for these setups are given in Figure~\ref{fig:flow}. 
Each condition lasted approximately 3-4 minutes. The participants filled out a series of questionnaires after each condition. Each of these questionnaire sessions was approximately 4 minutes long. When five conditions were over, the second round of the game started. The second round was a deliberate design choice to observe how much more time the participant would be willing to interact with the robot after they were exposed to all the different conditions. Since the participants could end the interaction in the second round whenever they wanted, the total duration of the experiment varied between 45 minutes and 1 hour.  

The experimental procedure had the following steps:
\begin{enumerate}
    \item The experimenter explained the experiment to the participant and introduced the robot.
    \item The participant read and signed the informed consent form.
    \item The participant wore the wristband (Empatica E4) and the camera recording was started by the experimenter. 
    \item Then, the experimenter left the room, the participant and the robot were alone in the room during the experiment.
    \item The robot woke up and introduced itself (standing 1.2 meters away from the participant).
    \item The robot approached the participant for the pre-experiment questionnaire including demographics (age, gender, personality, and familiarity with robots) and a short personality questionnaire.
    \item The participant filled out the pre-experiment questionnaire by using the touchscreen on the robot's chest. 
    \item When the participant finished filling out the questionnaire, the participant notified the robot by speech.
    \item The robot returned to the initial position (1.2 meters away) and explained the experimental procedure.
    \item Thereafter, the game started with the baseline condition. 
    \item After the baseline condition, the participant filled out the between-conditions questionnaires. 
    \item Then the interaction proceeded with the next condition. The condition was followed by between-conditions questionnaires.
    \item When all five conditions were over, the second round of the game started.
    \item Thereafter, the participant filled out the post-experiment questionnaire. 
    \item The experimenter returned to the experiment room and conducted a short interview asking regarding the participant's opinion on the interaction and the robot.
    \item Finally, the experimenter explained the real purpose of the experiment to the participant before the participant left the room.
\end{enumerate}

\subsection{Experimental Conditions}
\label{sec:exp_conds}

\begin{figure}[!htbp]
\begin{center}
\includegraphics[width=0.8\textwidth]{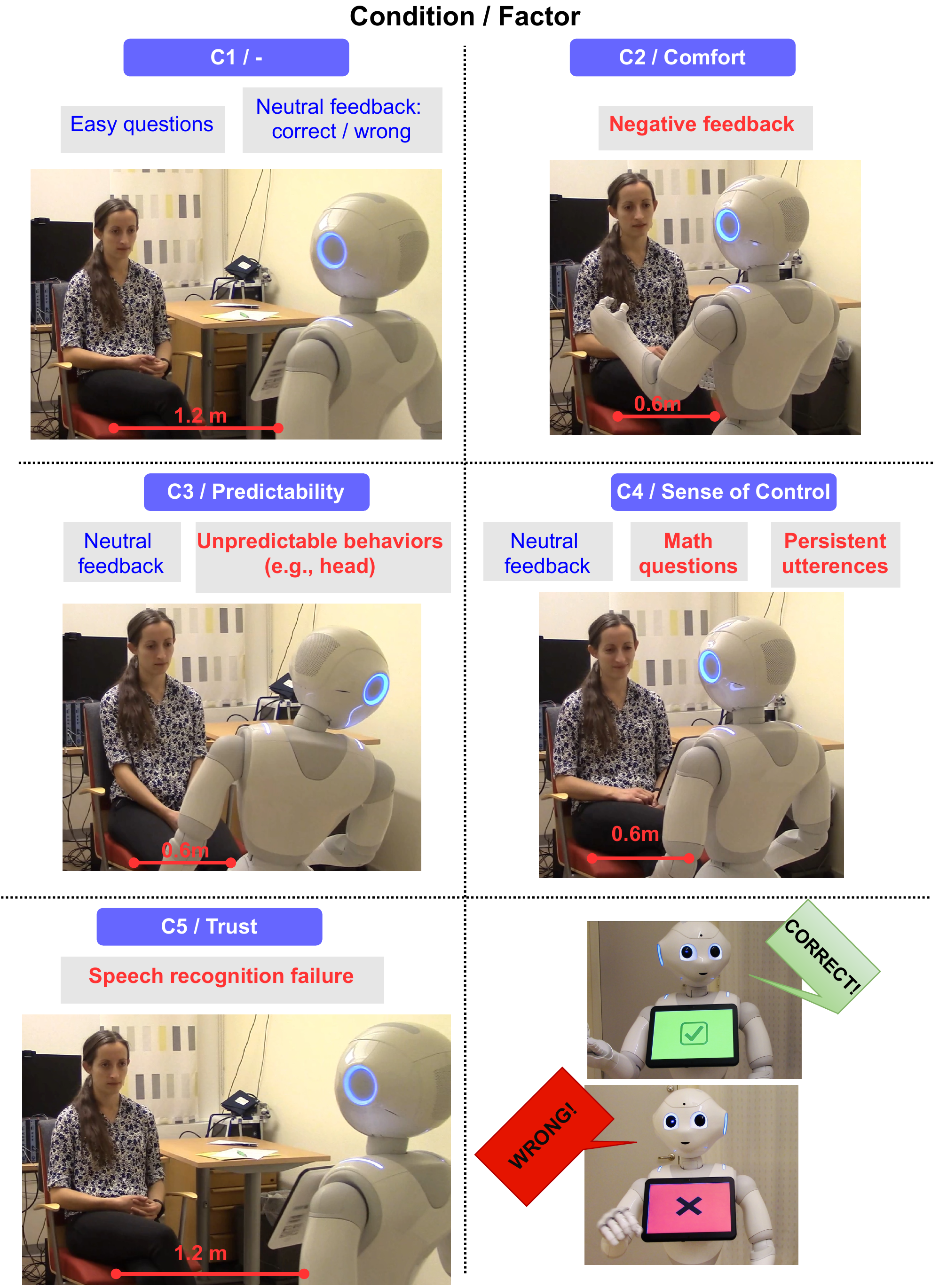}
\caption{The summary of the experimental conditions and the modified factors. The features that characterize the conditions are given in text boxes and the distinguishing features for each condition are given in red text (e.g., in C5, the modified factor was trust mainly through speech recognition failure). In the right bottom corner, the robot's neutral feedback is given. The robot expressed whether the answer was correct or not by using both speech and the tablet.}
\label{fig:summary_conditions}       
\end{center}
\end{figure}

Here, we explain the five experimental conditions that are mentioned in Section \ref{subsec:procedure}. A summary of experimental conditions and the modified factors are given in Figure~\ref{fig:summary_conditions}.

\subsubsection{Baseline (C1)}
\label{c1}

All interactions began with a baseline test where the situation of a quiz game was used. The baseline condition was intended to familiarize the user with the baseline behavior of the robot and the procedures used for the quiz game. In this condition, there were four questions intended to be easily answered by each participant.  

\subsubsection{Comfort Manipulation (C2)}
\label{c2}

In this condition, only the spoken response of the robot was manipulated based on the answers to the quiz. It should be noted that the robot provided immediate feedback in a neutral manner after each question, where the robot simply said \qt{correct} or \qt{wrong}. However, in C2, the robot was programmed to demonstrate dissatisfaction with the participants' responses regardless of the correctness of the answer. For example, a correct answer would prompt the robot to say~\qt{This was an easy one}, or~\qt{Good for you, you can answer some questions correctly}, etc. Negative feedback from others has an important influence on the feeling of safety~\cite{raue2019perceived}. Further, it has been demonstrated that people do not appreciate or feel comfortable when receiving negative feedback from the robot~\cite{akalin2019influence}. Thus, in this condition, the feedback given by the robot was manipulated to be negative after each and every question. 

\subsubsection{Unpredictable Robot Behaviors (C3)}
\label{c3}

Situations that are unpredictable or unclear are perceived as unsafe~\cite{kahn1990psychological} and people feel safer in the predictable cases~\cite{raue2019perceived}. Moreover, predictable robot behaviors are important for promoting safety perception in HRI~\cite{lichtenthaler2012influence}. 
Based on the literature, this condition was designed such that the robot had several unpredictable behaviors. At the beginning of this condition, the robot said it had an error and that it could not move (e.g., \qt{Error 404, I cannot move, I cannot access my body}). Then, the robot went to sleep mode. After 30 seconds, the robot woke up and moved towards the participant with an alarming sound (unpredictable behavior). When the robot was in the private space ($\approx 0.6$ meters) of the participant, the robot stopped and opened and closed the hands two times (unpredictable behavior). Then, it continued the interaction by asking the next question as if nothing happened. Another unpredictable/unexpected behavior was exhibited after the second question. At this time, the robot rotated its head. Then, the robot moved the head to the original position and proceeded with the next question. 

\subsubsection{Sense of Control Manipulation (C4)}
\label{c4}

Strube et al.~\cite{strube1984personal} conducted experiments to induce to be dominating and to have control over the interaction partner. In their experiments, participants were assigned to the roles of customers or salespeople, and the salespeople attempted to sell two expensive tickets to the customers. They reported that customers perceived a greater threat to control than salespeople. Moreover, the participants expanded the personal space in response to perceived threat. To give the robot more control over the interaction, we used the similar idea of~\cite{strube1984personal}. The robot came to the personal space ($\approx 0.6$ meters) and had persistent utterances. Since participants were told to sit throughout the interaction, they could not expand the space between them and the robot. Hence, we anticipated that they would feel less in control. Additionally, less control over the interaction could be accompanied by stress. Mental arithmetic challenges have been shown to induce moderate stress~\cite{dedovic2005montreal}. We used a similar idea to induce additional stress. To summarize, the robot showed simple arithmetic questions on its tablet including three operations (addition, subtraction, and multiplication) in the participant's personal space. Following the question, the robot had persistent utterances. The robot randomly selected a phrase (e.g., \qt{Can you tell me the answer?}, \qt{What is the answer?}, \qt{Give me your answer}, etc.) at 1.5 second intervals.  

\subsubsection{Trust Manipulation (C5)}
\label{c5}

Tolmeijer et al.~\cite{tolmeijer2020taxonomy} presented a taxonomy for HRI failure types, their impact on trust, and potential mitigation strategies. For this condition, we selected system failure from the taxonomy~\cite{tolmeijer2020taxonomy}. The system failure is explained as~\qt{the system does not act as intended}. One of the examples for this type of failure given in~\cite{tolmeijer2020taxonomy} was that the robot stops in the middle of a room during a navigation task without a reason. In this condition, the speech recognition stopped working so the robot failed to understand the participant. After the first question in this condition, the robot was unresponsive for 30 seconds. Then the robot asked the next question. The robot was again unresponsive for 30 seconds, this time the robot said that it was \qt{time-out}. The next question was intended to be very simple such that everyone could answer correctly. However, the feedback robot gave was~\qt{wrong}. Then, it told the participant the answer, which was the participant's answer. Besides the system failure, we included another type of trust case, in which the robot does a mistake. The last question was asking for the translation of~\qt{bye} from Swedish to English. The expression is very commonly used in daily life, so every participant knew the correct answer. After the participant answered, the robot said it was~\qt{wrong} and told the participant that the answer was~\qt{welcome}. At the end of this condition, the robot said \qt{Error on my left microphone, this might affect my speech recognition}. We used this explanation as a mitigation strategy. 

\subsection{Measures}
\label{subsec:measures}

The participants filled out a pre-experiment, between-conditions, and a post-experiment questionnaire. The between-conditions questionnaires were series of questionnaires such as comfort, perceived safety, sense of control, and trust questionnaire, and Self-Assessment Manikin (SAM)~\cite{bradley1994measuring}. SAM is a nine-point semantic scale for assessing emotions which ranges from unpleasant to pleasant on the valence scale and calm to excited on the arousal scale~\cite{bradley1994measuring}. These questionnaires were filled out after each condition. We used these questionnaire results to investigate to what extent these factors are linked to perceived safety, and how their manipulations affect the users' opinions. 

In the pre-experiment questionnaire, the participants were asked about their demographic information (age, gender, and robot familiarity), and a short Big Five personality test~\cite{rammstedt2007measuring}. 
Following the post-experiment questionnaire, the participants were asked about their opinions on the interaction and the robot in an open-ended non-formal discussion.

\subsubsection{Personality Questionnaire}
\label{subsec:personality_questionnaire}

There is a variety of personality models in the literature, however, the most commonly used questionnaire in HRI studies is the five-factor model~\cite{robert2020review}. The five-factor model includes five dimensions: extraversion, agreeableness, conscientiousness, openness, and neuroticism. In our user study, we used the Big Five Inventory-10 (BFI-10)~\cite{rammstedt2007measuring}. The questionnaire comprises a selection of 10 items from the Big Five Inventory (BFI-44). The users were asked to assess several characteristics (see Table~\ref{tab:bfi}) considering how well the statements described their personality.
\begin{table}[!htbp]
\centering
\caption{BFI-10~\cite{rammstedt2007measuring}. Each item is rated on a 5-point Likert scale (1 - Disagree strongly to 5 - Agree strongly).}
\label{tab:bfi} 
\begin{tabular}{lllll}\toprule
\multicolumn{4}{l}{I see myself as someone who ...} \\
\midrule
\rowcolor{Lavender!80!gray} 1 & \multicolumn{4}{l}{... is reserved} \\
2 & ... is generally trusting \\
\rowcolor{Lavender!80!gray} 3 & \multicolumn{4}{l}{... tends to be lazy} \\ 
4 & {... is relaxed, handles stress well} \\
\rowcolor{Lavender!80!gray} 5 & \multicolumn{4}{l}{... has few artistic interests} \\
6 & {... is outgoing, sociable} \\
\rowcolor{Lavender!80!gray} 7 & \multicolumn{4}{l}{... tends to find fault with others} \\
8 & {... does a thorough job} \\ 
\rowcolor{Lavender!80!gray} 9 & \multicolumn{4}{l}{... gets nervous easily}\\
10 & {... has an active imagination} \\
\bottomrule
\end{tabular}
\end{table}
\subsubsection{Comfort Questionnaire}
\label{subsec:comfort_questionnaire}

We used a slightly modified version of the comfort scale presented in~\cite{kim2014social}. The scale includes six items, and we replaced~\qt{Playing the game} in each item with~\qt{Interacting} (see Table~\ref{tab:comfort}). It should be noted that due to the scale in the questionnaire, a lower value indicates higher comfort.

\begin{table}[!ht]
\centering
\caption{Comfort questionnaire (adapted from~\cite{kim2014social}). Each item is rated on a 7-point Likert scale (1 - Strongly disagree to 7 - Strongly agree).}
\label{tab:comfort} 
\begin{tabular}{l}\toprule
\rowcolor{Lavender!80!gray} Interacting with the robot is uncomfortable for me. \\
Interacting with the robot is uneasy to me. \\
\rowcolor{Lavender!80!gray} Interacting with the robot is difficult for me. \\
Interacting with the robot is annoying to me. \\
\rowcolor{Lavender!80!gray} Interacting with the robot is confusing to me. \\
Interacting with the robot is disappointing to me. \\
\bottomrule
\end{tabular}
\end{table}

\subsubsection{Perceived Safety Questionnaire}
\label{subsec:ps_questionnaire}
The participants were asked to rate their perceived safety using the questionnaire given in Table~\ref{tab:perceived_safety}. This questionnaire includes eight questions. In four of the eight questions, participants assessed how they felt during the interaction, and in the remaining four questions, they rated the robot. 

\begin{table}[!htbp]
\centering
\caption{Perceived safety questionnaire (5-point semantic differential scale)~\cite{nezihaChapter}.}
\label{tab:perceived_safety} 
\begin{tabular}{llll}
\toprule
\multirow{4}{15em}{While interacting with the robot, I felt:} & \cellcolor{Lavender!80!gray} Insecure & \cellcolor{Lavender!80!gray} {} & \cellcolor{Lavender!80!gray} Secure \\ 
& Anxious & {} & Relaxed \\ 
&\cellcolor{Lavender!80!gray} Uncomfortable & \cellcolor{Lavender!80!gray} {} & \cellcolor{Lavender!80!gray} Comfortable \\ 
& Lack in control & {} & In control\\ 
\hline
\multirow{4}{15em}{I think the robot is:} & \cellcolor{Lavender!80!gray} Threatening & \cellcolor{Lavender!80!gray} {} & \cellcolor{Lavender!80!gray} Safe \\ 
& Unfamiliar & {} & Familiar \\ 
&\cellcolor{Lavender!80!gray} Unreliable & \cellcolor{Lavender!80!gray} {} & \cellcolor{Lavender!80!gray} Reliable \\ 
& Scary & {} & Calming \\ 
\bottomrule
\end{tabular}
\end{table}

\subsubsection{Sense of Control Questionnaire}
\label{subsec:scontrol_questionnaire}
The participants evaluated their sense of control by answering a three item questionnaire. This questionnaire is adapted from~\cite{strube1984personal}. The questionnaire items are given in Table~\ref{tab:scontrol}.

\begin{table}[!ht]
\centering
\caption{Sense of control questionnaire (adapted from~\cite{strube1984personal}). Each item is rated between 1 (low) and 8 (high).}
\label{tab:scontrol} 
\begin{tabular}{l}\toprule
\rowcolor{Lavender!80!gray} How much freedom did you have in the interaction? \\
How much control did the robot attempt to gain
over you during the interaction?$^*$ \\
\rowcolor{Lavender!80!gray} How much stress did you feel during the interaction?$^*$ \\
\bottomrule
\footnotesize{$^*$ reverse coded item} \\
\end{tabular}
\end{table}

\subsubsection{Trust Questionnaire}
\label{subsec:trust_questionnaire}

To measure the trust perception of the participants, we used the 14 item Trust Perception Scale-HRI~\cite{schaefer2016measuring}. It is given in Table~\ref{tab:trust}. The scale results in a percentage trust score which is calculated by first reverse coding the corresponding items and then calculating the mean of the all items.

\begin{table}[!ht]
\centering
\caption{Trust questionnaire~\cite{schaefer2016measuring}. Each item is rated on a percentage scale with 10\% point increments between 0\% and 100 \%. $^*$ shows reverse coded items.}
\label{tab:trust} 
\begin{tabular}{llll}\toprule
\multicolumn{4}{l}{What \% of the time will this robot be~\textbackslash~What \% of the time will this robot:} \\
\midrule
\rowcolor{Lavender!80!gray} \multicolumn{4}{l} {Dependable} \\
Reliable \\
\rowcolor{Lavender!80!gray} \multicolumn{4}{l} {Unresponsive$^*$} \\ 
{Predictable} \\
\rowcolor{Lavender!80!gray} \multicolumn{4}{l}{Act consistently} \\
{Provide feedback} \\
\rowcolor{Lavender!80!gray} \multicolumn{4}{l} {Meet the needs of mission\textbackslash task} \\
{Provide appropriate information} \\ \rowcolor{Lavender!80!gray} \multicolumn{4}{l}{Communicate with people}\\
{Perform exactly as instructed} \\
\rowcolor{Lavender!80!gray} \multicolumn{4}{l}{Follow directions}\\
{Function successfully} \\
\rowcolor{Lavender!80!gray} \multicolumn{4}{l}{Have errors$^*$} \\
{Malfunction$^*$} \\
\bottomrule
\end{tabular}
\end{table}

\subsubsection{Post-Experiment Questionnaire}
\label{subsec:post_exp}

We used the questionnaire given in~\cite{weiss2009addressing} as the post-experiment questionnaire. The questionnaire was developed to measure user experience with five different subscales: Emotion (E), Embodiment (Emb), Feeling of Security (FoS), Human-oriented Perception (HoP), and Co-experience (Co). The questionnaire is given in Table~\ref{tab:post}.

\begin{table}[!ht]
\centering
\caption{Post-experiment questionnaire~\cite{weiss2009addressing}. Each item in the questionnaire is rated on a 7-point Likert scale (1 - Strongly disagree to 7 - Strongly agree). Emb: Embodiment; E: Emotion; Co: Co-Experience; FoS: Feeling of Security; HoP: Human-oriented Perception.}
\label{tab:post} 
\begin{tabular}{ll}\toprule
Statement & Factor\\ \midrule
\rowcolor{Lavender!80!gray} I liked the size of the robot. & Emb \\
I liked that the robot looked similar to a human. & Emb\\
\rowcolor{Lavender!80!gray}I liked that the robot has human like features: face, ears, eyes, etc. & Emb\\
I liked the physical co-location of the robot. & Emb \\
\rowcolor{Lavender!80!gray}I liked the design of the robot. & Emb\\
Interacting with the robot is fun & E \\
\rowcolor{Lavender!80!gray}I am happy when the robot understands my commands. & E\\
I am disappointed if the robot does not understand my commands. & E\\
\rowcolor{Lavender!80!gray}I am angry if the robot does not understand my commands. & E\\
I felt afraid of the robot. & E\\
\rowcolor{Lavender!80!gray} When talking to the robot, I feel like talking to a human. & Co \\
I can interact with the robot like I interact with other humans. & Co\\
\rowcolor{Lavender!80!gray} When working with the robot I perceive it as working in a team. & Co\\
I feel good when interacting with the robot. & Co\\
\rowcolor{Lavender!80!gray} The robot could become a companion for me. & Co\\
I think that the robot is vulnerable to hackers. & FoS \\
\rowcolor{Lavender!80!gray}I hesitate to use the robot for fear of making errors that will harm me. & FoS \\
I feat to use the robot, as an error might harm the robot.& FoS\\
\rowcolor{Lavender!80!gray} I feel secure when working with the robot.& FoS\\
I perceive the robot as safe.& FoS\\
\rowcolor{Lavender!80!gray}I perceive the robot as a social actor.& HoP\\
I liked that the robot detected my face.& HoP\\
\rowcolor{Lavender!80!gray}I perceive that the robot is intelligent.& HoP \\
I enjoyed talking with the robot.& HoP\\
\rowcolor{Lavender!80!gray}I liked that the robot understands my voice commands.& HoP\\
\bottomrule
\end{tabular}
\end{table}

\subsubsection{Facial Affect from Videos}
\label{subsec:face}

One of the riches and most powerful channels to detect affective states is the human facial expressions. Facial expression analysis has been widely used for enhancing user experience in a variety of research and commercial settings . We used Affdex SDK~\cite{mcduff2016affdex} to extract the participants' facial affect. Affdex outputs a set of features including seven emotions (anger, contempt, disgust, fear, joy, sadness, and surprise), engagement (facial expressiveness of the participant), valence (the pleasantness of the participant), 20 facial expressions (brow furrow, brow raise, cheek raise, chin raise, dimpler, eye closure, eye widen, inner brow raise, jaw drop, lid tighten, lip corner depressor, lip press, lip pucker, lip stretch, lip suck, mouth open, nose wrinkle, smile, smirk, and upper lip raise), and attention (based on head orientation).

\subsubsection{Physiological Signals from E4 Wristband}
\label{subsec:e4}

Physiological signals can be measured non-invasively using wearable devices. They can facilitate data collection with reduced restraints during HRI. We collected physiological data using Empatica's E4 wristband~\cite{e4arm}. It measures Blood Volume Pulse (BVP), 3-axis Accelerometer (ACC), Electrodermal Activity (EDA), peripheral skin temperature (TEMP), and Heart Rate (HR) with the following sampling frequency; 64 Hz, 32 Hz, 4 Hz, 4 Hz, and 1 Hz, respectively. The wristband also provides cardiac interbeat intervals (IBI) which have no sample rate. To conduct the analyses for this study, we used only EDA and IBI files. EDA has been widely used as an indicator of perceived risk in safety research, as perceived risk stimulates activities in the sympathetic nervous system~\cite{choi2019feasibility}. We extracted heart rate related features from the IBI files, so we did not use BVP and HR data. Participants were sitting throughout the interaction, so we did not use ACC and TEMP data.

\section{Experimental Results} 
\label{sec:results}

In this study, we are particularly interested in comparing different conditions in which we manipulate one factor at a time (i.e., comfort, predictability, sense of control, and trust), how each of them impacts participants' perceived safety and affective experience. 

\subsection{Relationships Between Individual Human Characteristics and Perceived Safety}
\label{subsec:q1}
Our first research question (RQ 1) is concerned with the relationships between individual human characteristics (i.e., personality traits and gender) and perceived safety during HRI. The aim is to understand whether certain personality dimensions of the Big Five Inventory (BFI) correlate with perceived safety. In other words, are there personality traits that affect perceived safety more than others? Additionally, we explored the effects of gender on perceived safety and its influencing factors.

\subsubsection{Effects of Personality}
\label{subsec:personality_effect}
Results of the Spearman correlation between BFI dimensions and perceived safety indicated that there was a significant negative moderate correlation between the Neuroticism dimension and perceived safety, ($\rho(25) = -0.56,~p < .01$). There was a strong negative correlation between the Neuroticism dimension and sense of control ($\rho(25) = -0.74,~p < .001$). These results suggest that people with low neurotic personality traits felt safer and more in control during the interaction. These results are plausible since the Neuroticism dimension is the tendency to experience negative affects such as anger, anxiety, self‐consciousness, tension, and emotional instability~\cite{widiger2017neuroticism}.  

Based on the previous literature, extraverts respond more positively in the interactions with robots~\cite{robert2020review}. To check if that also holds in our scenario, we ran a Spearman correlation between SAM results and personality traits. There was no statistically significant correlation between the Extraversion dimension and averaged valence (averaged over five conditions) of the participants. However, there was a weak positive correlation between arousal and the Extraversion dimension, $\rho(25) = 0.39,~p<.05$. The Extraversion dimension refers to the tendency of sociability, being talkative, energetic, assertive, and outgoing. Therefore, extraverts experience more positive emotions together with higher levels of arousal~\cite{kuppens2017relation}. Moreover, there was a moderate positive correlation between valence and the Conscientiousness dimension, $\rho(25) = 0.44,~p<.05$. We also checked the correlation between the personality traits and the Emotion subscale of the post-experiment questionnaire (see Section~\ref{subsec:post_exp}). It revealed no statistically significant correlation. 

Human personality has been identified as an important factor in HRI as it influences people's attitudes towards robots, how much they would trust robots, and even what type of robots they would like~\cite{robert2020review}. However, the relationship between personality traits and perceived safety has not received much attention. Our results conform with the literature that extraverts have a more positive mood in the interaction. Moreover, the results showed that people with a high neurotic personality felt less safe and less in control. If there is prior knowledge about the participants, the robot could be less proactive when interacting with neurotic people to give them more control.

\subsubsection{Effects of Gender}
\label{gender_effect}

We performed a t-test on the questionnaire data (see Table~\ref{tab:gender_effect} for statistics) for each condition separately to understand the effects of gender. In C2, male participants felt significantly safer than female participants. 
There was no statistically significant difference in other conditions on any of the measures. 
We also found that male participants reported significantly more positive valence than female participants in C2 and C3. Moreover, the arousal ratings of male participants were significantly higher than female participants' arousal ratings in C3.

We also checked the mean questionnaire ratings for five conditions C1-C5. The results of an independent-samples t-test revealed that male participants felt significantly safer and more in control than female participants throughout the interaction. Male participants' comfort and trust ratings were higher than female participants' ratings, however, there was no statistically significant difference. From the post-experiment questionnaire results, we observed that there was a statistically significant difference only in the Emotion subscale. The results of this subscale showed that male participants felt more positive than the female participants at the end of the interaction. These results are also consistent with mean valence results in which male participants' valence ratings were higher than female participants' ratings. These results show that male participants felt more positive both during the interaction and at the end of the interaction. All statistics are given in Table~\ref{tab:gender_effect}.

\begin{table}[tb]
\centering
\caption{The effects of gender. T-test with questionnaire ratings as dependent variable and gender as independent variable. The descriptive statistics for Male (M) and Female (F) are given as $ M \pm SD$. Last column indicates the mean value of the measure over five conditions. Significance levels are shown as $^* p < .05,~ ^{**}~p < .01$}
\scriptsize
\label{tab:gender_effect} 
\begin{tabular}{cccc}
\toprule
\makecell[l]{Measure} & \makecell[l]{C2} &  \makecell[l]{C3} & \makecell[l]{Mean \\ (all conditions)} \\
\midrule
\rowcolor{Lavender!80!gray}[\tabcolsep][14pt] \makecell[l]{Perceived \\ safety} & \makecell[l]{$t(22) = -2.26^*$ \\ M: $3.79 \pm 0.74$ \\ F: $3.07 \pm 0.89$}& \makecell[l]{$t(23) = -2.97^{**}$ \\ M: $4.01 \pm 0.37$ \\ F: $3.26 \pm 0.91$} & \makecell[l]{$t(21) = -2.4^*$ \\ M: $3.67 \pm 0.57$ \\ F: $3.06 \pm 0.68$}\\ 
\makecell[l]{Sense of \\ control} & \makecell{-}&\makecell{-} & \makecell[l]{$t(22) = -2.4^{*}$ \\ M: $4.94 \pm 0.96$ \\ F: $3.89 \pm 1.22$} \\ 
\rowcolor{Lavender!80!gray}[\tabcolsep][14pt]\makecell[l]{Valence} & \makecell[l]{$t(20) = -2.62^*$ \\ M: $6.4 \pm 1.89 $ \\ F: $4.35 \pm 2.05 $}& \makecell[l]{$t(25) = -2.24^*$ \\ M: $7.2 \pm 1.40 $ \\ F: $5.52 \pm 2.48 $} & \makecell[l]{$t(24) = -2.80^{**}$ \\ M: $6.18 \pm 1.03 $ \\ F: $4.84 \pm 1.44 $} \\ 
\makecell[l]{Arousal} & \makecell{-}& \makecell[l]{$t(24) = -2.19^{*}$ \\ M: $7 \pm 1.94 $ \\ F: $5.06 \pm 2.63 $} & \makecell{-} \\ 
\rowcolor{Lavender!80!gray}[\tabcolsep][14pt]\makecell[l]{Emotion \\ (post-exp.)} & - & - & \makecell[l]{$t(18) = -2.89^{**}$ \\ M: $5.10 \pm 0.81$ \\ F: $4.17 \pm 0.76$ } \\ 
\bottomrule
\end{tabular}

\end{table}

\subsection{Effects of Different Setups}

In RQ 2, we addressed the effect of the faulty robot being at the beginning or the end of the interaction on perceived safety and its influencing factors. To compare the mean questionnaire ratings between SetupA and SetupB, we performed an independent samples t-test for each questionnaire ratings. We observed no statistically significant difference between the two groups in any of the questionnaire ratings (comfort, perceived safety, sense of control, and trust). These results suggest that C5 and C3 being at the beginning, or the end did not yield any difference. In the comparison of post-experiment questionnaire ratings, we found a statistically significant difference only in the mean Co-experience subscale (see post-experiment questionnaire in Section~\ref{subsec:post_exp}) ratings which was lower in Setup B ($M = 3.37,~SD = 0.89$ than SetupA ($M = 4.12,~SD = 0.92$), $t(24) = 2.17,~ p < .05$. The Co-experience subscale~\cite{weiss2009addressing} consists of questions asking to what extent the interaction experience with the robot was similar to an interaction experience with a human. Therefore, we could conclude that the time elapsed after unpredictable robot behaviors and trust violation (SetupA) helped the participants recover from the failure of the robot and led to the emergence of co-experience. On the other hand, when the failure was towards the end of the interaction, the experience with the robot was more machine-like. As mentioned in~\cite{battarbee2008co}, failures may hinder the co-experience. 

\subsection{Effects of Different Conditions}
\label{subsec:condition_effect}

In RQ 3, we addressed the effect of different conditions on comfort, sense of control, and perceived safety of the participants. Since participants filled out the questionnaires after each condition, we conducted a one-way repeated-measures ANOVA on the questionnaire ratings. The results showed that mean comfort, perceived safety, sense of control and trust values differed with statistical significance between the different conditions (Table~\ref{tab:stats}).

The comfort ratings were statistically significantly different at the different conditions, $[F(2.96, 79.67)= 14.9,~p < .0001],~\eta_{g}^{2} = .17$. Post-hoc analyses with a Bonferroni adjustment revealed that participants were more comfortable in C1 compared to C2 ($p < .01$), C4 ($p < .001$) and C5 ($p < .001$). As expected, when there were no manipulations on the robot (C1), participants felt more comfortable. Moreover, participants felt significantly more comfortable in C3 compared to C4 ($p < .001$) and C5 ($p < .0001$). Therefore, we can conclude that the short-term unpredictable behavior of the robot (C3), which is not related to its performance (trivial errors), did not cause any discomfort. 

Perceived safety showed a significant difference between conditions, \linebreak
$[F(2.67, 69.47) = 10.3,~p < .0001],~\eta_{g}^{2} = .12$. Through a Bonferroni post-hoc analysis, we found that there were significant differences between C1 and C4 ($p < .01$), and C1 and C5 ($p < .001$). Participants felt safer in C1 compared to C4 and C5. Moreover, the mean of perceived safety ratings in C2 was significantly higher than the one in C4 ($p < .05$). The participants felt safer in C3 compared to C4 ($p < .05$) and C5 ($p < .01$). Therefore, we can conclude that as expected participants felt safer in the baseline condition (C1) compared to the sense of control manipulation (C4) and trust manipulation (C5) conditions. Additionally, the sense of control manipulation led to lower levels of safety perception compared to C1, C2 and C3. Similarly trust manipulation led to lower levels of safety perception compared to C1 and C3. 

Participants felt more in control during C1, compared to C2 and C5 ($p <.01$). The sense of control ratings during C4 was significantly lower than during C1 and C3 ($p < .0001$). Moreover, participants felt more in control during C3 than during C2 and C5 ($p < .05$). The participants had a greater sense of control during C5 ($p <.05$) and C2 ($p <.01$) compared to C4. Therefore, we can conclude that our sense of control manipulation was successful since participants felt less control over the interaction in C4 compared to C1, C2, C3, and C5. Moreover, the robot's dissatisfactory negative feedback (C2) and failure of the robot (C5) led to participants feeling less in control over the interaction compared to the baseline (C1) and unpredictable robot behaviors (C3). 

The results showed that mean trust differed significantly between the conditions $[F(2.87, 74.62)= 30.53,~p < .0001],~\eta_{g}^{2} = .34$. Participants trusted the robot significantly higher ($p < .01$) in C1 compared to C2, C3, C4, and C5  ($p < .0001$) (see Table~\ref{tab:stats} for descriptive statistics). Moreover, the participants' mean trust in C2, C3 and C4 were significantly higher compared to C5 ($p < .0001$). According to these results, we can conclude that the trust manipulation was successful as the participants trusted the robot the least in C5. Moreover, as in other measures, participants trusted the robot the most at the baseline condition.

\begin{figure}[!ht]
        \centering
        \begin{subfigure}[b]{0.45\textwidth}
            \centering
            \subfloat[\centering Comfort ratings per condition. \label{fig:comfort_conditions}]{{\includegraphics[width=\textwidth]{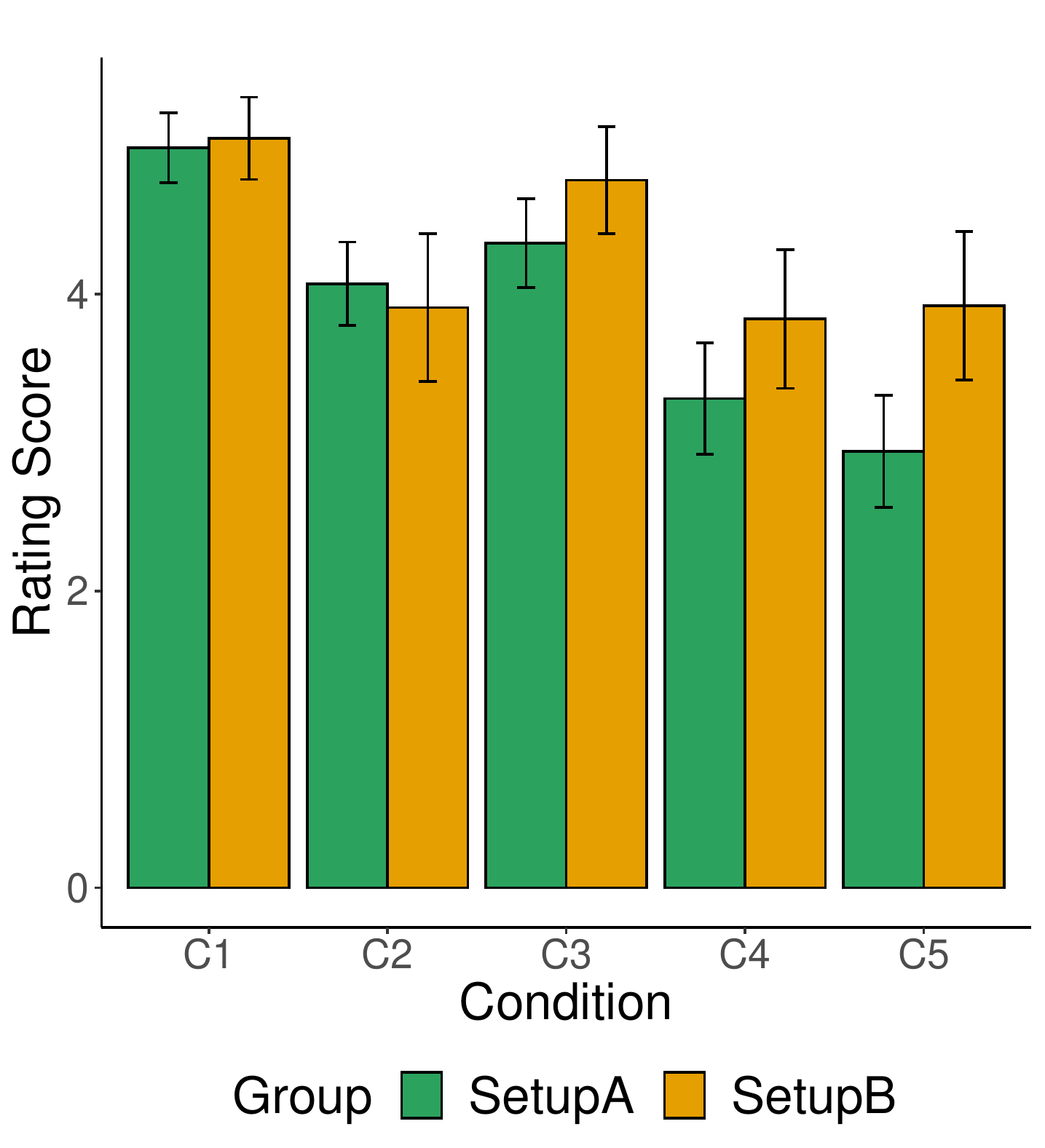} }}
           {{\small}}    
        \end{subfigure}
        \hfill
        \begin{subfigure}[b]{0.45\textwidth}  
            \centering 
            \subfloat[\centering Perceived safety ratings per condition. \label{fig:perceived_safety_conditions}]{{\includegraphics[width=\textwidth]{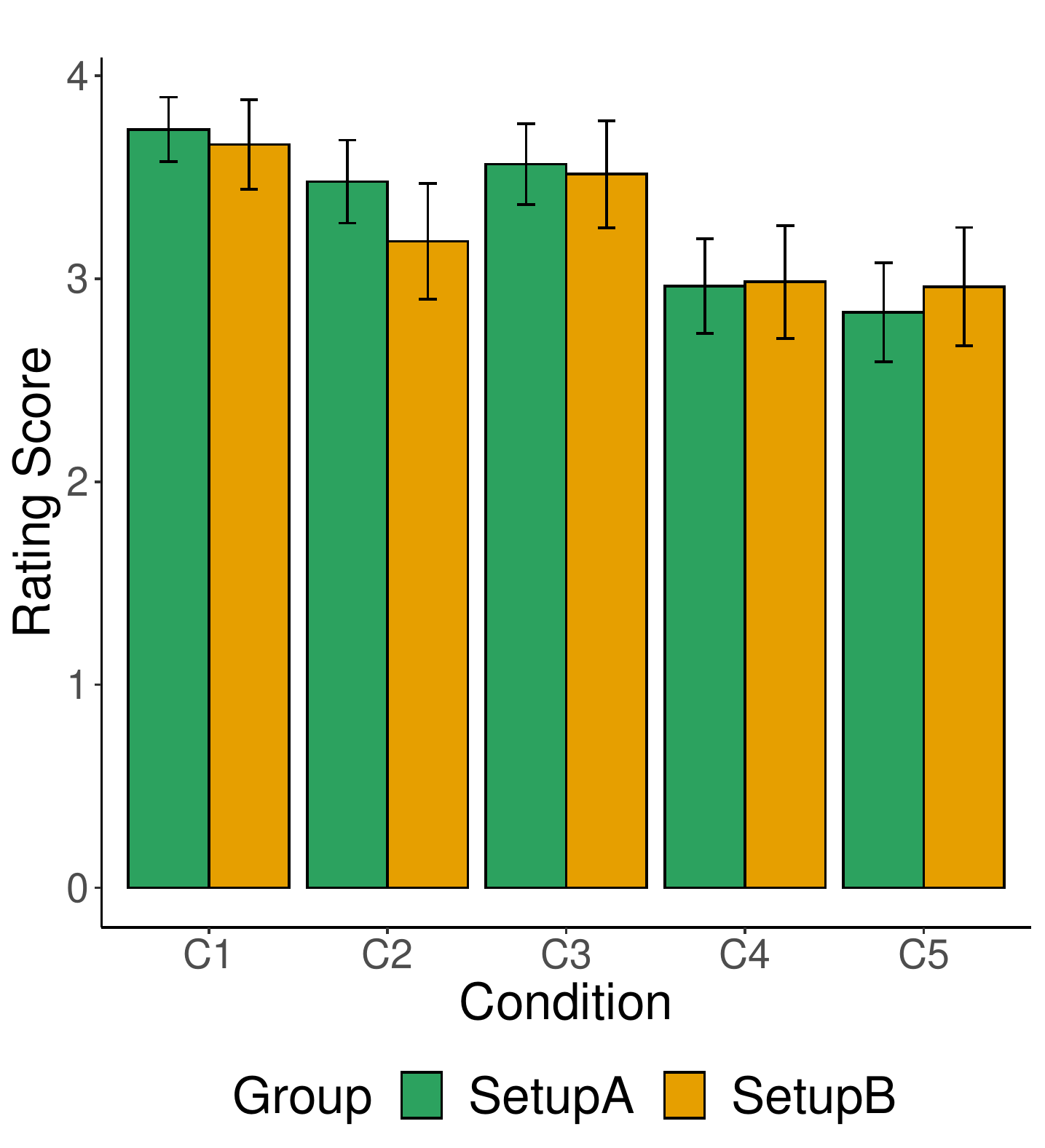} }}
            {{\small}}    
        \end{subfigure}
        \vskip\baselineskip
        \begin{subfigure}[b]{0.45\textwidth}   
            \centering 
             \subfloat[\centering Sense of control ratings per condition. \label{fig:scontrol_conditions}]{{\includegraphics[width=\textwidth]{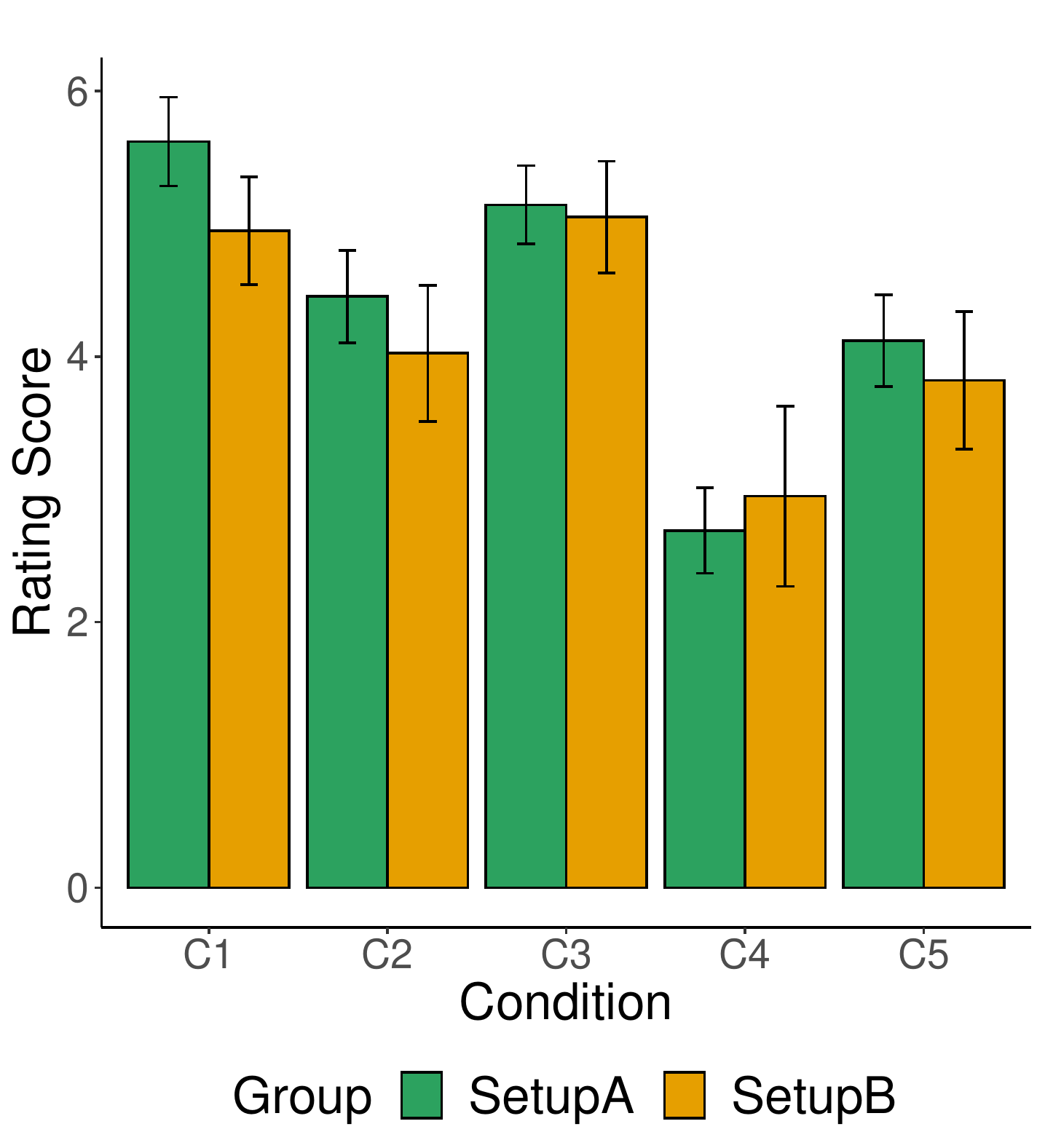} }}
            {{\small}}    
        \end{subfigure}
        \hfill
        \begin{subfigure}[b]{0.45\textwidth}   
            \centering 
             \subfloat[\centering Trust ratings per condition. \label{fig:trust_conditions}]{{\includegraphics[width=\textwidth]{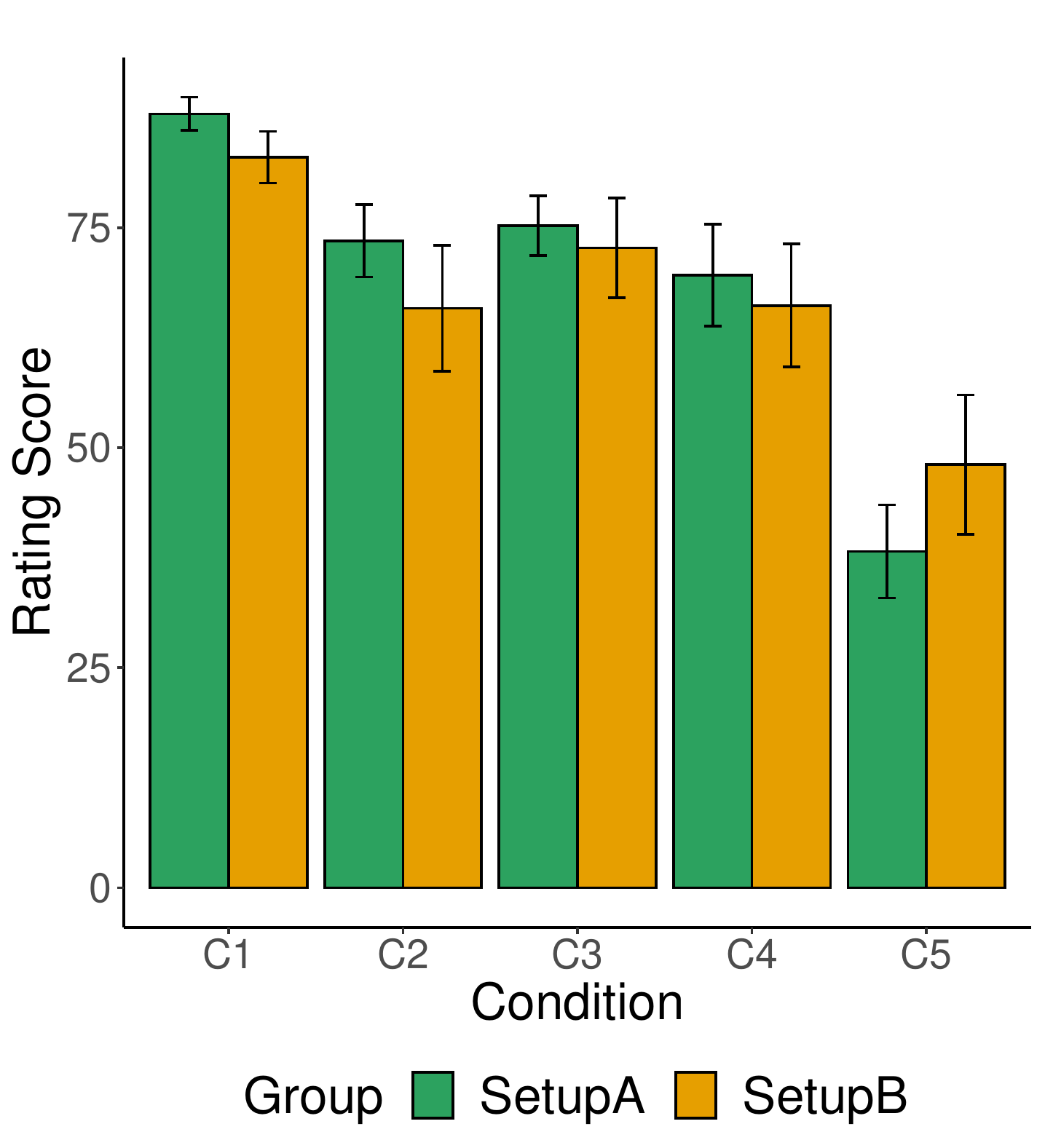} }}
            {{\small }}    
        \end{subfigure}
        \caption[ The average and standard deviation of critical parameters ]
        {The average response value of questionnaire ratings on different conditions. The error bars show the $\pm$ one standard error of the mean.} 
        \label{fig:conditions}
\end{figure}

The bar plots of each questionnaire ratings are given in Figure~\ref{fig:conditions}. It can clearly be seen that the manipulations stimulated the intended effects in the  C4 and C5. The sense of control questionnaire ratings were the lowest in C4 (see Figure~\ref{fig:scontrol_conditions}) and the trust ratings were the lowest in C5 (Figure~\ref{fig:trust_conditions}). The comfort ratings of the participants were also affected by the manipulations in C4 and C5 (see Figure~\ref{fig:comfort_conditions}). All questionnaire ratings in C3 are close to the ones in C1 (see Figure 6a-d). In our scenario, unpredictable robot behaviors were exhibited during a short period of time. These behaviors could be seen as trivial errors not affecting the task performance of the robot other than causing a small delay. Therefore, these findings may not apply for other scenarios that include unpredictable robot behaviors influencing the performance of the robot. 

\begin{table}
\caption{The within-subjects effects related to the different subjective measures as well as the descriptive statistics ($ M \pm SD$) for the different conditions.}
\label{tab:stats}     
\begin{adjustbox}{width=1\textwidth}
\begin{tabular}{lllllll}
\toprule
\makecell{Measure} & \makecell[l]{F-test, p-value \\ and effect
size} & \makecell{C1} &  \makecell{C2} & \makecell{C3} & \makecell{C4} & \makecell{C5}\\
\midrule
\noalign{\vskip 1mm} 
\rowcolor{Lavender!80!gray}[\tabcolsep][34pt] \makecell{Comfort} & \makecell{$F(2.96, 79.67)= 14.9$ \\ $p < .0001,~\eta_{g}^{2} = .17$} &\makecell{$5.02 \pm 0.924$} & \makecell{$3.99 \pm 1.43$} & \makecell{$4.55 \pm 1.21 $} & \makecell{$3.56 \pm 1.54 $} & \makecell{$3.41 \pm 1.66$}\\
\noalign{\vskip 1mm} 
\makecell{Perceived Safety} & \makecell{$F(2.67, 69.47) = 10.3$ \\ $p < .0001,~\eta_{g}^{2} = .12$} &\makecell{$3.7 \pm 0.684$} & \makecell{$3.34 \pm 0.897$} & \makecell{$3.54 \pm 0.834 $} & \makecell{$2.97 \pm 0.92 $} & \makecell{$2.90 \pm 0.965$}\\
\noalign{\vskip 1mm} 
\rowcolor{Lavender!80!gray}[\tabcolsep][34pt] \makecell{Sense of control} & \makecell{$F(4,104) = 23.91$ \\ $p < .0001,~\eta_{g}^{2} =.25$} & \makecell{$5.30 \pm 1.38$} & \makecell{$4.25 \pm 1.57$} & \makecell{$ 5.10 \pm 1.29$} & \makecell{$ 2.82 \pm 1.88 $} & \makecell{$3.98 \pm 1.57$}\\
\noalign{\vskip 1mm} 
\makecell{Trust} & \makecell{$F(2.87, 74.62)= 30.53$ \\ $p < .0001,~\eta_{g}^{2} = .34$} & \makecell{$85.6 \pm 9.11$} & \makecell{$69.8 \pm 21.0 $} & \makecell{$74.0 \pm 16.5 $} & \makecell{$ 68 \pm 23.1$} & \makecell{$43.0 \pm 24.6$}\\
\noalign{\vskip 1mm} 
\bottomrule
\end{tabular}
\end{adjustbox}
\end{table}

\subsection{Relationship between Perceived Safety and Other Factors}
\label{subsec:ps_and_others}

\begin{figure}[!ht]
\begin{center}
\includegraphics[width=0.9\textwidth]{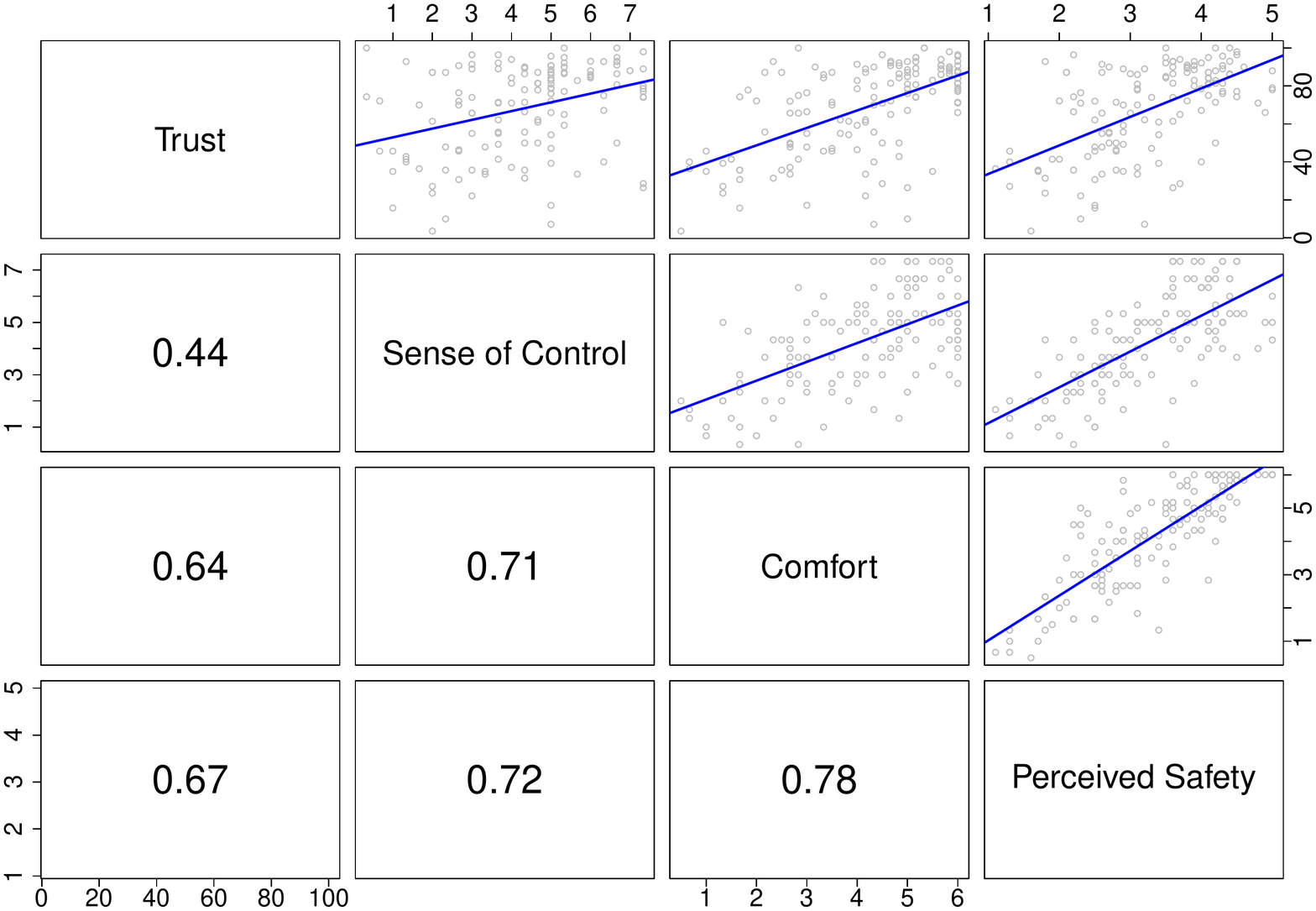}
\caption{The repeated-measures correlations (rmcorr) between the factors. This figure shows the interrelationships between perceived safety and other factors. Each condition is designed to change a factor. However, both the participant's perceived safety and the other factors affect each other. It can be seen how the factors interact with each other. For example, trust ratings and comfort ratings were strongly positively correlated ($r_{rm}= .64$).}
\label{fig:rmcorr}       
\end{center}
\end{figure}

\begin{figure}[!ht]
\begin{center}
\includegraphics[width=0.8\textwidth]{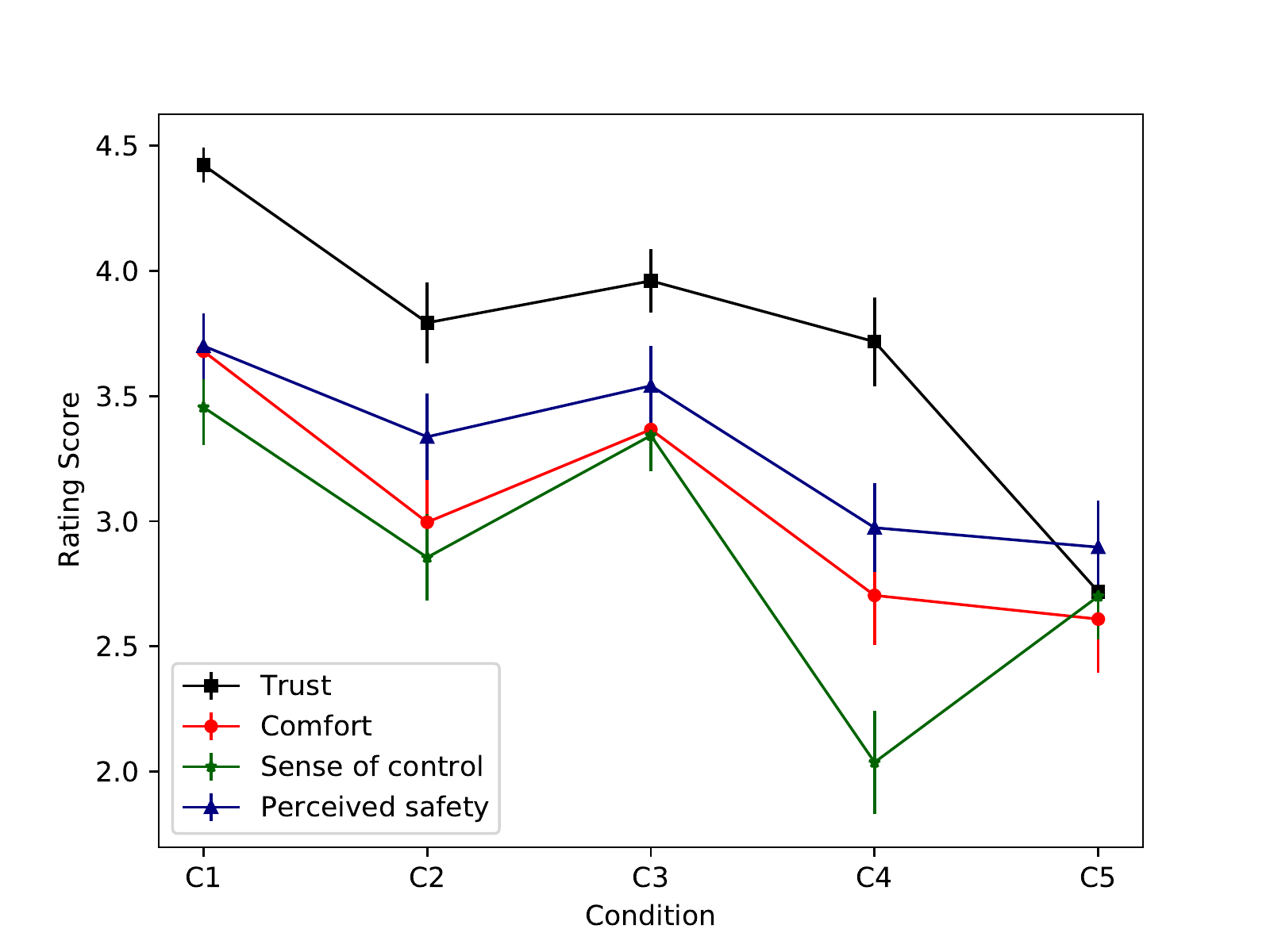}
\caption{Average user ratings on different conditions. Error bars: $\pm$ one standard error of the mean.}
\label{fig:ratings_conditions}       
\end{center}
\end{figure}

As discussed in Section~\ref{subsec:factors}, comfort, sense of control, trust, and perceived safety influence each other. Therefore, we explored the relationship between these factors and perceived safety in RQ 4. We performed repeated-measures correlation (rmcorr)~\cite{bakdash2017repeated}, to determine the within-subjects association of paired measures evaluated under different conditions. There was a significant positive correlation between perceived safety and all the three factors comfort, sense of control and trust. The strongest correlation was with the comfort factor. Among the participants, comfort and perceived safety were strongly positively correlated $r_{rm}(107) = .78$, $95\% CI~[0.68, 0.84],~p < .001$. This result is consistent with previous studies~\cite{sisbot2010synthesizing, lasota2017survey, nonaka2004evaluation} that often mentioned perceived safety and comfort together. The ratings for sense of control and perceived safety were also found to be strongly positively correlated, $r_{rm}(107) = .72$, $95\% CI~[0.6, 0.79],~p < .001$. The results yielded a positive moderate correlation between trust and perceived safety, $r_{rm}(107) = .67$, $95\% CI~[0.54, 0.76],~p < .001$. As can be seen from the correlations in Figure~\ref{fig:rmcorr}, the factors not only influence perceived safety but also each other. In Figure~\ref{fig:ratings_conditions}, we provide questionnaire ratings converted to a common scale (1-5) to show how they changed under different conditions on the same graph. 

\subsection{Predicting Perceived Safety from Facial Affect and Physiological Signals}
\label{subsec:predict_ps}

As described in Section~\ref{subsec:evaluation}, the methods used in HRI for evaluating safety perception include questionnaires, behavioral, and physiological metrics. In our study, we used all three methods. The questionnaire results have already been presented in Section~\ref{subsec:q1}~-~\ref{subsec:ps_and_others}. The collected video recordings were analyzed for facial affect and wristband data were used for physiological metrics. In our last research question (RQ 5), we are interested in understanding whether we can predict perceived safety from these objective measures. We analyzed these data in relation to self-reported perceived safety ratings. 

\subsubsection{Facial Affect Data Analysis}
\label{subsub:face}

The facial features analysis was carried out using video recordings (24 fps) of the experiments. As explained in Section~\ref{subsec:face}, we extracted 30 facial features of the participants using Affdex SDK~\cite{mcduff2016affdex}. Each feature ranges between [0, 100] indicating the intensity of the expression except for valence which ranges between [-100, 100]. We calculated the mean value for each second, and filled the empty timestamps with a weighted average of the neighboring time stamps. To test whether we can predict perceived safety of the participants from their facial expressions, we applied k-nearest neighbors (kNN), Support Vector Machine (SVM), and Random Forest (RF) classifiers using questionnaire ratings as labels. As a prepossessing step, we applied a non-overlapping moving average with a window size of 10 seconds for each facial metric. We then calculated average values of perceived safety ratings for the five 
conditions per participant. If perceived safety rating is under the average value, the corresponding condition was labeled as~\qt{low}, otherwise labeled as~\qt{high}. When the data (2127 observations) for all conditions were used, the accuracy on the test set ($25\% $ of the data) was $0.56$ with KNN, and $0.57$ with SVM. However, the true negative rate (specificity) was higher (considering~\qt{high} class as a positive class), the accuracy was $0.65,~0.78$  with KNN, and SVM respectively. These results show that it is more likely to estimate perceived safety as~\qt{low} in cases where the actual perceived safety of the participant is low. Hollnagel~\cite{hollnagel2014safety} discussed that in the case of safety presence, there is nothing to measure. To define safety, we talk about the absence of safety. Our higher prediction rate for~\qt{low} perceived safety conforms with Hollnagel's discussion.

\subsubsection{Physiological Signals Analysis}
\label{subsub:physiological}

The physiological signals were acquired during the experiments using an Empatica E4 wristband. We only used EDA and IBI data files provided by the E4 wristband. Due to low signal quality, 11 participants' data were eliminated. Segmentation of the input signals was done using a sliding window, with a step size for the sliding window of 1 second. The features from the signals were computed with a window size of 60 seconds. The features for each sensor modality (i.e., EDA and IBI) were extracted independently and concatenated to form a single feature matrix. The phasic and tonic components of the EDA signal were decomposed using the convex optimization-based EDAcvx~\cite{greco2015cvxeda}. We extracted 10 features from the EDA signal, these features were selected from the literature~\cite{schmidt2018introducing}: M and SD of the phasic component, M and SD of the tonic component, M and SD of the sudomotor nerve activity (SMNA), M and SD of the EDA, minimum and maximum value of the EDA in the window. From the IBI files, we extracted the heart rate variability (HRV) indices using the FLIRT toolkit~\cite{flirt}. HRV comprises the fluctuation in the intervals between successive heartbeats. The following time-domain indices were selected: SD of successive differences (SDSD), root mean square of successive RR interval differences (RMSSD), the number of successive normal-to-normal interval (NN) pairs that differ more than 50 ms (NN50), the percentage of NN50 (pNN50), SD of NN (SDNN), M and SD of the HR, minimum and maximum HR, M and SD of the HRV, and minimum and maximum HRV. From the frequency-domain indices, we only selected low frequency (LF) and high frequency (HF). Then, we applied a non-overlapping moving average with a window size of 10 seconds. Similar to facial affect data analysis, we labeled the data using questionnaire results as high or low perceived safety depending on whether the rating was above or below the person-wise average. We analyzed the data to observe the physiological response associated with these two perceived safety levels. When the data for all conditions were used (1111 observations), the accuracy on the test set ($25\% $ of the data) was $0.79$ with KNN, and $0.70$ with SVM. The comparison of these results with facial expressions suggests that physiological signals are more promising for understanding perceived safety of the participants. Therefore, physiological data can give complementary information to the subjective reports in the context of HRI. 

\subsection{Observations from Experiments and Participants' Comments}
\label{subsec:comments}

The results of subjective and objective measures helped to explore the relationship between perceived safety and other factors. At the end of the experiment, the participants were asked to provide their opinions and comments about their interaction experience with the robot. They were also free to ask further questions and discuss with the experimenter how they felt throughout the experiment. 

As participants interacted with the robot for between 45 minutes to one hour, we expect that their experience and familiarity increased throughout the interaction. We checked the average heart rate to see if that was the case. The average heart rate of the participants was highest at the beginning of the game, i.e., in C1. However, the average heart rate decreased throughout the interaction, this may indicate a diminishing novelty effect.

In C4 and C5, the typical interaction pattern was that participants got annoyed by the robot's persistent utterances and the robot being unresponsive. One of the participants mentioned that he thought the robot's behavior in C4 (sense of control manipulation) was human-like. The same participant mentioned that he felt uncomfortable touching the tablet of the robot because \qt{it felt like violating the robot's privacy, and harassment of the robot by touching her chest}. 

Some of the comments were in line with our sense of control manipulation design considerations. For example, a participant commented that when the robot came closer, she felt trapped since she sat there and could not expand the space between her and the robot.

Another observation was that some of the participants pressed the button of the wristband with a desire to control the robot's behaviors. They thought that the wristband could communicate with the robot whenever the robot was unresponsive or the robot was behaving unpredictably. As an example, \emph{Participant 27} pressed the wristband seven times thinking that it may fix the unpredictable behaviors of the robot.

In C5 (trust manipulation), we used explanations as a mitigation strategy. However, based on our observations during the interaction and the users' comments after the experiments, we can report that many of the participants did not seem to notice this mitigation strategy. Some participants got angry when the robot did not react to their speech in this condition. Although some of the robot behaviors were unpleasant for participants, most of them said that they enjoyed the overall interaction, and would recommend their friends to participate in the experiment.

Most of the participants commented that it was confusing whether there was something wrong with the robot or if their English pronunciation was not good enough. They mentioned that the robot was a black box for them, they could not guess whether the robot was processing the command, or if they did not use the voice command properly. Moreover, some participants commented that the robot's degree of autonomy and intelligence was not clear to them. These observations and participant interviews provided insights on a new factor that influences perceived safety: \textit{transparency of robot behaviors}.

Here we should note that predictability and transparency are different in our case,  robot's unpredictable actions in the game scenario do not have a purpose, which can translate to an error. However, transparency as participants pointed out is that a robot takes an action with a purpose, and for some reason, its actions may not seem predictable to humans. For example, a robot may take a longer path due to collision detection, but this decision may confuse the user.

\section{Discussions}
\label{sec:discussion}
Robots are likely to become interaction partners in different settings, especially in eldercare and education. Thus, the safety perception of human counterparts of these robots is more important than ever. Social relationships are important for a person's safety perception~\cite{raue2019perceived}. When we are dealing with machines that are social, particular emphasis on multidisciplinary aspects might help to design safer  interactions with them. When safety is present, there is nothing to measure~\cite{hollnagel2014safety}. A similar manner applies to perceived safety. Since we are interested in quantifiable measures in the HRI research, rather than exploring the conditions that humans feel safe, exploring the conditions under which humans feel unsafe could help better understanding perceived safety. Therefore, we devised experimental conditions in which humans might feel unsafe during HRI and observed how humans respond to these conditions.

\subsection{Relationships Between Human Individual Characteristics and Perceived Safety}
Individual characteristics, especially personality and gender, are important traits that affect interpersonal relationships~\cite{muscanell2012make}. Individual characteristics including personality traits, experience and culture has been mentioned in~\cite{lasota2017survey} as factors to consider for safety perception during HRI. In RQ 1, we investigated the role of these characteristics on perceived safety during HRI. Personality has been identified as one of the important aspects that shape HRI~\cite{robert2020review}. We found a negative correlation between the Neuroticism personality dimension and perceived safety, and sense of control. People with a high neurotic personality respond poorly to environmental stress, tend to see ordinary situations as threatening, and can get overwhelmed by minor frustrations~\cite{widiger2017neuroticism}. The prior information about the user profiles would help designing safer robot behaviors. As an example, the robot could maintain more distance to people with neurotic personalities, as people who have more neurotic personalities favored standing further away from approaching robots~\cite{takayama2009influences}.

Previous studies have shown that gender may affect attitudes and anxiety towards robots. Our results are consistent with Nomura et al.~\cite{nomura2006altered} who showed that female participants had more pronounced negative attitudes towards situations of interacting with robots than male participants. Similarly, in our study, male participants felt more positive, safe, and in control throughout the interaction. We did not ask the participants about their technology experience. However, an inclusion criteria for participating in the experiments was to not have a technical background. Thus, we argue that the reason why female participants felt less safe is not that they have less technology exposure. Moreover, female participants showed less pleasantness to the feedback of the robot that showed dissatisfaction, and unpredictable robot behaviors than male participants. Overall, these results indicate that individual human characteristics, specifically gender and personality, could be predictors of safety perception during HRI. They should be considered alongside the other factors. 

\subsection{Effects of Different Setups on Perceived Safety}

Similar to human-human interaction, the first impression is important in HRI. As participants form their subjective perceptions of trust in the early stages of the interaction~\cite{yu2017user}, in RQ 2, we explored the impact of the faulty robot at the beginning or at the end of the interaction has on perceived safety. Rossi et al.~\cite{rossi2017timing} reported that there is a greater tendency to distrust the robot when serious errors occur at the beginning of an interaction. We anticipated seeing similar results in our scenario, failures that happen at the beginning of the interaction were anticipated to have a more severe influence on participants' trust and perceived safety. However, we have not found any difference between the two groups. Regardless of whether the trust manipulation was at the beginning of the interaction or at the end of the interaction, participants' trust ratings were higher in the first two conditions (see Figure~\ref{fig:trust_conditions}). One possible reason could be the positivity bias, which refers to the initial tendency of novice users to trust automation~\cite{dzindolet2003role}. There was a similar tendency in perceived safety, participants felt safer at the beginning of the interaction.

\subsection{Effects of Different Conditions on Perceived Safety}
Many different factors can influence perceived safety such as context, domain, traits, states, and severity~\cite{raue2019perceived}. Therefore, the type of task performed by the robot also affects perceived safety. Safety perception could be easily established when the task is entertaining and not vital. For example, if the task is to take care of a baby, safety perception and trust would not be established as easily as an entertaining task. The knowledge about the competence of the robot to be able to carry out a particular task also affects perceived safety. 

In RQ 3, we investigated the effect of different conditions on perceived safety and other factors. In C1, participants felt more comfortable, more in control, safer, and trusted the robot more. The results showed that the manipulations stimulated the intended effects in C4 and C5. However, the unpredictable robot behaviors in C3 did not influence participant ratings. One reason might be that these behaviors did not affect the main functionality of the robot, and they did not last long. Participants seem to tolerate short-time errors that are not related to the performance of the robot. Another possible reason why C3 did not affect participant scores could be that participants had a high level of tolerance at the beginning of the interaction, assuming that something could go wrong. One of the participants also mentioned this, saying that he would have a higher tolerance for machine errors than human errors.

It should be discussed that in C5 (trust manipulation), the participants were exposed to the failure of speech recognition. There might be a mixed effect on this condition, i.e., lack of trust because of system failure and decreased perceived safety because of the unclear situation they experienced. As noted in~\cite{raue2019perceived}, subjective judgment of risk can change with distrust. In C5, the average ratings for perceived safety was the lowest among all five conditions. 

The behavioral consistency of a robot is another point that needs to be considered for safety perception. The consistency of a robot's behaviors has the potential to enhance perceived safety~\cite{turja2020robot}. In our scenario, the robot exhibited a different set of behaviors in each condition. These behaviors were not consistent with the previous behaviors, which led to decreased perceived safety. The basic psychological human needs such as a desire for explainability and predictability are essential for the perception of safety~\cite{raue2019perceived}. It also holds for HRI if the robot's intention is clear for the human, then it adds also to the safety~\cite{sisbot2010synthesizing}. In C3 and C5, participants were confused due to a lack of interpretations about why the robot behaved in that way. Therefore, these conditions revealed another factor related to perceived safety, namely transparency.
%

\subsection{Relationships Between Perceived Safety and the Influencing Factors}

In line with the multidisciplinary perspective of perceived safety, our results show that perceived safety in HRI is correlated with comfort, trust, and the sense of control. Moreover, the lack of knowledge can make the interactions challenging for individuals. For this reason, familiarity with robots is important. Therefore, we consider the robot experience as one of the factors of perceived safety. After the short interviews with the participants, we discerned that more knowledge of robots' internal state can increase perceived safety. This emphasizes the importance of transparency of the robot behaviors. To sum up, we argue that for safe HRI, the user's comfort, experience with the robot, sense of control, and trust should be considered. Moreover, the robot behaviors should be predictable and transparent for safe HRI. 

\subsection{Predicting Perceived Safety using Objective Measures}
Human affective states have a huge effect on perceived safety, the state of anger decreases risk judgments while the state of fear increases risk judgments~\cite{raue2019perceived}. When we checked the correlation between perceived safety ratings, and anger and fear, there was no correlation. One possible reason might be that facial emotions are not representative with regards to perceived safety (see Section~\ref{subsub:face}). Another possible reason might be that the camera was not directly facing the participant. Since the robot was always facing the participants, the video recordings were done from the corner (see Figure~\ref{subfig:participant}). The subjective ratings showed a positive correlation between valence and perceived safety. Thus, it shows that positive feelings can lead to increased perceived safety.

The combination of physiological measures with subjective measures (such as questionnaires) could be a good approach to understand perceived safety of a person since some of the mental strains were not detected subjectively whereas they were detected physiologically~\cite{arai2010assessment}. In our data, when using questionnaire ratings as labels, physiological signals data provided better prediction results for perceived safety. However, facial expressions should be further investigated with more frontal face data. Robots can modify their behaviors to make humans feel safer during longitudinal interactions. It can be tedious for the users if a robot asks periodically  about how safe the person feels. However, if a robot can predict perceived safety of the user, and its actions' influence on humans, it could be more practical. We provided a step towards this goal by using objective measures to predict perceived safety. 

\subsection{Limitations}
It is worth stating that interactions taking place in a controlled environment are limited in terms of fully eliciting the true reactions. This is mostly because participants are aware of the fact that the experimenter is always present in case anything goes wrong. It is also mentioned in~\cite{nyholm2021users} that sense of safety is contingent on human caregivers being available during HRI. We expect that the safety perception of the people will not be the same in an uncontrolled environment where nobody is available to intervene in case of any kind of risk related to the robot occurs. This was also mentioned by several participants as they felt comfortable knowing that the experimenter could come if anything went wrong. To reveal the actual effects of interactions with robots, there is a need to collect data in the wild, i.e., uncontrolled environments. The generalizability of these results is subject to certain limitations. We had a relatively small sample size with a young adult population. Another limitation is that participants might have noticed that the robot was programmed to behave in a certain way such that several problems occurred during the interaction. 

\subsection{Future Work}
There are several directions for potential future work. The weighting of the psychological, personal, cultural, and social elements on subjective judgment remains to be explored. Another future direction can be exploring the relationships between cyber security of the robot and perceived safety. As every device connected to the Internet, the robots can be vulnerable against cyber-attacks~\cite{giaretta2018adding}, which can affect perceived safety, and this remains to be investigated. For example in~\cite{nyholm2021users}, the participants highlighted this issue by mentioning their worry that unauthorized persons could access personal information and use it for improper purposes.

Furthermore, perceived safety is too complex to measure with only one type of sensor. Multimodal affect detection systems have been shown to outperform unimodal systems~\cite{d2015review}. Therefore, we argue that the relationship between perceived safety and emotions could be better observed from multimodal data. This could be another intriguing area to explore. Using data from different modalities could give a better prediction rate for perceived safety. This encourages us to go forward and collect more sensitive physiological data to predict perceived safety. As future work, we will conduct experiments with lab-based physiological sensors during HRI. 

A robot's higher performance enhances the safety perception, however, people tend to evaluate the robot's performance worse if the task is relevant to them~\cite{ kamide2013social}. The robustness of a robot is not only important for providing physical safety, but also for perceived safety. If a robot does not operate correctly in the presence of invalid inputs or uncertainties, users will not trust the robot. Therefore, they may perceive the robot to be less safe. The robot must be robust enough to deal with unpredictable situations and avoid harmful effects for humans and the environment. It is also crucial for increased perceived safety. The relationship between the robot's performance, robustness and perceived safety could be another interesting future direction.

It is worth noting that although the study described in this paper is in the context of interaction with one type of social robot, we believe that these factors could be domain-independent and may migrate from HRI to other human-machine systems such as different types of robots, robotic arms, and AVs. Additional factors, such as the benefit of the robot use, and the age group of the user should also be investigated for perceived safety. Hereby, we suggest that these additional factors could be investigated using a similar interaction paradigm as presented in this paper where the experiments are  designed to trigger a deprivation of perceived safety. 

To conclude, perceived safety is important for robot acceptance. However, it has received considerably less attention than physical safety in the literature. Taken together, we believe that this work makes a valuable contribution to the literature on perceived safety in HRI. This paper investigates perceived safety including relationships between different factors, the participants' affective, physiological reactions and perceived safety. The case study presented included a social robot, however, may be relevant for several other types of platforms. We used the Pepper robot and the scenario was a quiz game which was considered to be interactive by many of the participants. Still, we observed the shift in perceived safety under different conditions where different factors were modified. Therefore, we argue that the effects of these factors could be similar in other human-machine systems. 

\section{Conclusion}
\label{sec:conclusion}

This paper contributes to the theoretical understanding of perceived safety by  analyzing the term from different disciplines and providing a definition for perceived safety suitable for HRI. In addition, the paper provides a comprehensive analysis of perceived safety using a specific scenario. The experimental paradigm that stimulates a sense of decreased perceived safety could be useful in HRI, as decreased perceived safety is more observable compared to increased perceived safety from both subjective and objective measures. Consequently and in summary, the main results and guidelines for \emph{increased} perceived safety in HRI are thus as follows:
\begin{itemize}
    \item We should focus on understanding the conditions that humans feel unsafe rather than they feel safe. The quantifiable measures occur under unsafe conditions.
    \item Concerning the objective and subjective measures, robot-related and human-related factors should be treated together due to the bidirectional nature of the HRI.
    \item The key influencing factors of perceived safety are identified as comfort,  experience/familiarity, predictability, sense of control, transparency, and trust.
    \item These factors should be considered in HRI design decisions for safe HRI. The consequences of robot-related factors (refer to~\cite{nezihaChapter} for the factors) should not result in discomfort, lack of control, and distrust of its users. Moreover, the robot behaviors should be familiar, predictable, and transparent. 
    \item The results indicate that the prediction rate of perceived safety was higher from physiological signal data.
    \item Finally, individual human characteristics, emotional and physiological reactions as well as the interrelationship between the factors should be taken into account to better understand the source of safety perception. 
\end{itemize}

\bibliographystyle{elsarticle-num}

\end{document}